Discussion Paper: Theory of opinion distribution in human relations where trust and distrust mixed(2020)

# Convex Regions of Opinion Dynamics, Approaches to the Complexity of Binary Consensus with Reference to Addiction and Obliviousness:Integrated Dimer Model Perspective

Yasuko Kawahata †

Faculty of Sociology, Department of Media Sociology, Rikkyo University, 3-34-1 Nishi-Ikebukuro,Toshima-ku, Tokyo, 171-8501, JAPAN.
ykawahata@rikkyo.ac.jp,kawahata.lab3@damp.tottori-u.ac.jp

**Abstract:** The field of opinion dynamics has grown out of early research that applied magnetic physics methods to better understand social opinion formation. A central challenge in this field is to model how diverse opinions coexist and influence each other. Opinions are rarely in complete agreement on social issues, and the formation of biaxial consensus is also very rare. To address this issue, Ishii and Kawahata (2018) proposed an extended version of the bounded trust model that takes into account the influence of dissent, distrust, and mass media. Their model aims to mimic more realistic social opinion dynamics by introducing coefficients representing the degree of trust and distrust, rather than assuming convergence of opinions. In this study, we apply the dimer construct and the dimer model to develop a theoretical framework. Through numerical simulations, we also show how the proposed model applies to actual social opinion formation. The study employs a model that computes the castellain matrix $K$, the distribution function $Z$, and the probability of dimer configuration $P(D)$ for convex regions with different positions and distances, and analyzes how changes in convex regions affect the probability of dimer configuration. The model takes into account "dependence" and "forgetting" in opinion formation and discusses "distance" and "location" of opinions. Numerical simulation results show how the proposed model can capture real-world social opinion formation processes. This study provides a foundation for a deeper understanding of the social opinion formation process and the development of effective strategies for real-world social problems. Overall, this research provides a new methodology for capturing and understanding the complex dynamics of opinion formation in the field of opinion dynamics. This theoretical framework and numerical simulation-based approach provides new perspectives in the fields of social science, physics, and computational modeling, and aims to provide new perspectives and contribute to a deeper understanding of social opinion formation.

## 1. Introduction

The field of opinion dynamics has grown steadily since the early studies that applied magnetic physics methods to better understand social opinion formation. In particular, it is rare for people's opinions to be in complete agreement, and in many cases, especially on social issues, biaxial consensus is very rare. A key challenge in this area is to understand and model how diverse opinions coexist and influence each other.

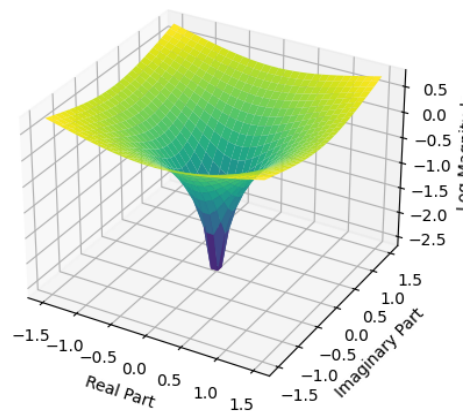

Copyright © Yasuko Kawahata

Fig. 1: Energy function based on a convex domain in the amoeba complement $E(x, e)$

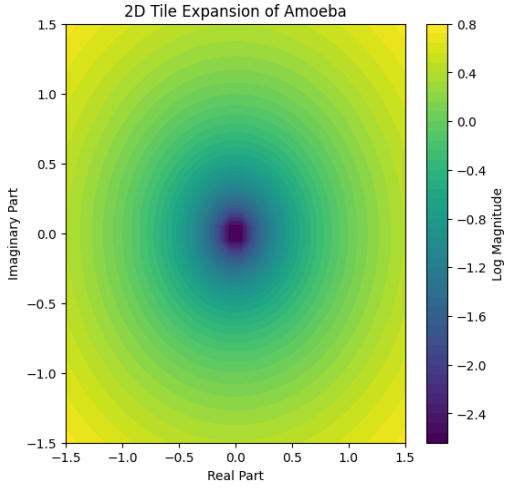

Fig. 2: 2D Tile Expansion of Amoeba: Energy function based on a convex domain in the amoeba complement $E(x, e)$

To address this challenge, Ishii and Kawahata (2018) proposed an extended version of the bounded credibility model that introduces new parameters for dissent, distrust, and mass media influence. Their model aims to capture more realistic social opinion dynamics by introducing coefficients representing the degree of trust and distrust, rather than assuming convergence of opinions. This approach sheds new light on the ways in which individuals are influenced by social opinion.

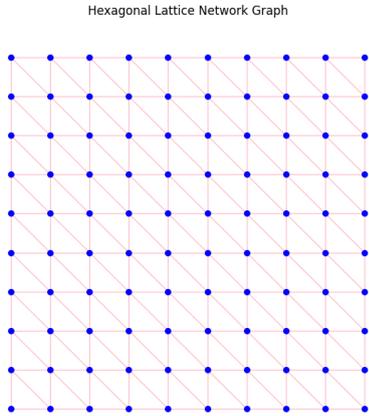

Fig. 3: Results, Hexagonal Lattice Network Graph

Based on this trust-distrust model (TDM), this paper proposes a new approach to opinion dynamics that applies the dimer allocation and Ising model. The purpose of this study is to explore how the interaction between trust and distrust affects social opinion formation. In particular, we use mathematical models to analyze how various external stimuli, such as mass media, third-party opinions, and economic and political factors, affect people's opinions. Our approach is to mathematically represent the dynamics of trust and distrust, which have not been addressed by traditional models. This theoretical framework provides new insights into the polarization of opinions, the process of consensus building, and how these are reflected in social behavior. In this paper, we apply the dimer configuration and dimer model to develop the theoretical framework, and use numerical simulations to show how the proposed model applies to actual social opinion formation. This research aims to contribute to a deeper understanding of social opinion formation by providing new perspectives in the fields of social science, physics, and computational modeling. In particular, this paper employs a model that computes the castellain matrix $K$, the distribution function $Z$, and the probability of dimer placement $P(D)$ for a set of convex regions with different positions and distances. This introduced a method to analyze how changes in convex regions affect the probability of dimer placement, modeling that takes into account "dependence" or "forgetting" of the distribution of opinions in opinion formation, as well as "distance" and "location" to the opinions. We also attempted to analyze from the perspective of the z-axis in order to discuss the "distance" and "positional relationship" to opinions. The energy function $E(x, e)$ is set based on the convex region in the complement of the amoeba that describes the convex region. Introduce the idea of assigning lower energy values to edges within the convex region and higher energy values to edges outside the convex region. We introduce an algorithm that calculates the castellane matrix based on the convex region and then calculates the probability of dimer placement.

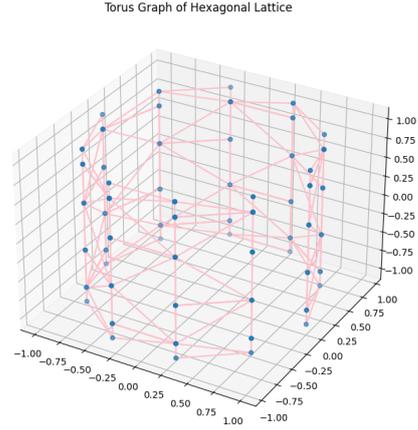

Fig. 4: Results, Surface Representation:Torus Graph of Hexagonal Lattice

This theoretical framework provides an important extension to existing models in the field of opinion dynamics. In traditional models, the distribution of opinions is usually modeled under simplifying assumptions. For example, opinions are assumed to vary only within a certain range or only certain influencing factors are considered. However, the ac-tual social opinion formation process is much more complex, with many different factors interacting. The proposed model incorporates this complexity and allows for a more realistic simulation of opinion formation.

Furthermore, this study will explore in greater depth the role of individual influencing factors in opinion formation. In particular, it will focus on how external influences such as mass media, third-party opinions, and economic and political conditions affect individual opinions. This will provide a more detailed understanding of the importance of external stimuli and their dynamics in the opinion formation process. The proposed model also provides a new way to integrate diversity and conflict in opinions and new insights into social consensus and disagreement.

Numerical simulation results demonstrate how the proposed model can capture real-world social opinion formation processes. These simulations are an important step in validating the model and exploring its applicability in real social phenomena. Simulations allow us to observe how individual opinions change and evolve over time. This provides a basis for better understanding the process of social opinion formation and for developing effective strategies for real social problems.

Overall, this research provides a new methodology for capturing and understanding the complex dynamics of opinion formation in the field of opinion dynamics. This theoretical framework and numerical simulation-based approach aims to provide new perspectives in the fields of social science, physics, and computational modeling to contribute to a deeper understanding of social opinion formation.

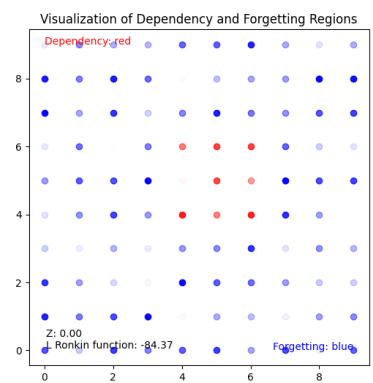

Fig. 5: Results, Dependency and Forgetting Regions Matrix

## 2. Previous Studies

The model developed in this paper is a modeling of opinions that depend on certain conditions, such as location, various opinions, etc., "attachment phenomena" and "forgetting phenomena" as their reversible phenomena. Therefore, we will first summarize some examples of research on dependencies that are likely to occur in recent years, such as research on digital dependencies. Then, we summarize previous studies on the theory used in this paper.

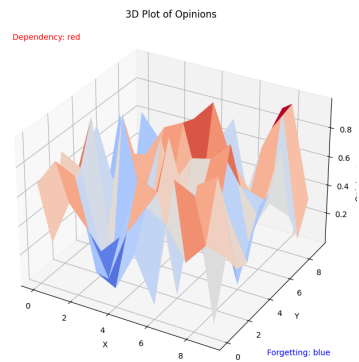

Fig. 6: Results, 3D Dependency and Forgetting Regions Kasteleyn Matrix

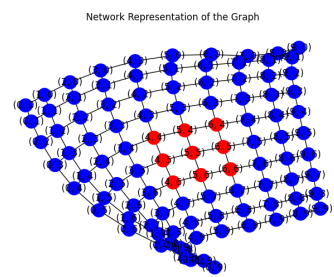

Fig. 7: Results, Network Representation of the Graph:Dependency and Forgetting Regions

### 2.1 Research on Digital Dependence

#### 2.1.1 Research Case Studies on the Dependency Scale

First, a paper published by Young in 1998 (Young, K. S., 1998) addresses the topic of "Internet Dependence: the Emergence of a New Clinical Disorder." In this paper, the concept of Internet addiction is proposed and initial considerations regarding its characteristics and diagnosis are presented.

Next, a paper published by Griffiths in 2005 (Griffiths, M., 2005) focuses on the topic of "An 'elemental' model of addiction within the biopsychosocial framework." This paper proposed an "elemental" model for understanding addiction within a biopsychosocial framework, bringing a new perspective to the understanding of addiction.

Also, a paper published by Kuss, Griffiths, Karila, and Billieux in 2014 (Kuss, D. J., Griffiths, M. D., Karila, L., Billieux, J., 2014), entitled "Internet dependence: epidemiology over the past decade A Systematic Review of Research" focuses on the topic. In this study, a systematic review of epidemiological studies on Internet dependence over the past decade is provided.

In addition, a paper published by Young in 1998 (Young, K. S., 1998) focuses on the topic "Caught in the Net: recognizing the signs of Internet dependence and winning strategies for recovery." The paper provides information on strategies for recognizing and recovering from the signs of Internet addiction.

Finally, a paper published by Kim, Davis, Hedayati, and Kao in 2009 (Kim, H. K., Davis, K. E., Hedayati, A., Kao, Y. C., 2009), entitled "Internet Addiction in Turkish Adolescents: Problematic Internet Use reliability and validity of the scale". The study provides research findings on the reliability and validity of Internet addiction among Turkish adolescents.

### 2.1.2 Smartphone Use and Challenges

This paper published by Koob and Volkow in 2016 (Koob, G. F., and Volkow, N. D., 2016) addresses the topic of "The Neurobiology of Addiction by Analyzing Central Neural Circuits". This study focused on the neurobiological aspects of addiction and detailed the role of central circuits.

Next, a paper published by Grant et al. in 2010 (Grant, J. E., Potenza, M. N., Weinstein, A., and Gorelick, D. A., 2010) focuses on the topic of "Introduction to Behavioral Addiction". This study provides an overview of behavioral addiction and its importance.

Additionally, a paper published by Witkiewitz and Marlatt in 2004 (Witkiewitz, K., and Marlatt, G. A., 2004) focuses on the topic "Relapse Prevention of Alcohol and Drug Problems: It Was Zen, This is the Way". In this study, approaches to relapse prevention for alcohol and drug problems are explored and the metaphors of Zen and the Way are used.

Another paper published by Griffiths in 2005 (Griffiths, M. D., 2005) focuses on the theme of "'component' models of addiction within a biopsychosocial framework. This study proposes a "component" model that views addiction within a framework that takes into account biopsychosocial factors.

Finally, a paper published by Leshner in 1997 (Leshner, A. I., 1997) focuses on the theme that "addiction is a brain disease and it matters." In this paper, addiction is a brain disease and it matters. This case study will be used to organize with respect to case studies on dependence on smartphones and other digital environments. And this paper published by Roberts and David in 2016 (Roberts, J. A., and David, M. E., 2016) focuses on "Partner Fabbing and Romantic Partner Relationship Satisfaction." This study investigated the impact of smartphone use on romantic relationships and found the impact of smartphone use between partners on relationship satisfaction.

Second, a paper published by Elhai et al. in 2017 (Elhai, J. D., Levine, J. C., Dvorak, R. D., and Hall, B. J., 2017) found that "feelings of FOMO (missing something without fear), need for touch, anxiety, and depression are problematic smartphone associated with problematic smartphone use" are explored in this study. The study examines the association between problematic smartphone use and psychological health, showing that factors such as FOMO are associated with problematic smartphone use.

Additionally, a paper published by Tavakolizadeh et al. in 2019 (Tavakolizadeh, J., Ataran, M., and Aminian, M., 2019), entitled "Problematic Smartphone Use: Conceptual Overview and Relationship to Anxiety and Depression Psychopathology. A Systematic Review" focused on the topic. The study presents a comprehensive review of the association between problematic smartphone use and anxiety symptoms and depression.

Another paper published by Liu et al. in 2020 (Liu, C. H., Lin, S. H., Pan, Y. C., and Lin, Y. H., 2020) addresses the topic of "Smartphone gaming and frequent usage patterns are associated with smartphone dependence." In this study, the relationship between smartphone game use and addiction was investigated, with particular emphasis on the association with frequent usage patterns.

Finally, a paper published by Lee and Lee (Lee, S. Y., and Lee, D., 2021) focuses on the topic of "The role of loneliness and motivation to belong in smartphone dependence among youth: a chained intermediate model." In this study, the effects of loneliness and willingness to belong on smartphone dependence are revealed through a chained, intermediate model.

### 2.1.3 Social Networking Service (SNS) Dependence

First, a paper published by Kuss and Griffiths in 2017 (Kuss, D. J., and Griffiths, M. D., 2017) addresses the topic of "Social Networking Sites and Addiction: 10 Lessons Learned". The study summarized 10 key lessons learned about addiction related to social networking site use.

Next, a paper published by Andreassen et al. in 2012 (Andreassen, C. S., Torsheim, T., Brunborg, G. S., and Pallesen, S., 2012) focuses on the topic "Development of the Facebook Dependence Scale". This study involved the development and validation of the Facebook Addiction Scale, a scale to measure addiction related to Facebook use.

In addition, a paper published in 2013 by Kwon et al. (Kwon, M., Lee, J. Y., Won, W. Y., Park, J. W., Min, J. A., Hahn, C., ... and Kim, D. J., 2013) focuses on the topic of "Development and Validation of the Smartphone Dependency Scale (SAS)." In this study, the Smartphone Addiction Scale (SAS) was developed as a scale to measure addiction related to smartphone use.

Another paper published by Ryan et al. in 2014 (Ryan, T., Chester, A., Reece, J., and Xenos, S., 2014) focused on the topic "Facebook use and abuse: a review of Facebook dependence". The study reviews addictions associated with Facebook use and provides insights into its use and abuse.

Finally, a paper published in 2017 by Bányai et al.

(Bányai, F., Zsila, Á., Király, O., Maraz, A., Elekes, Z., Griffiths, M. D., ... and Demetrovics, Z., 2017) focuses on the topic "Problematic Social Media Use: Results from a Large Nationally Representative Sample of Youth." The study reports findings from a nationally representative survey of youth on problematic social media use.

### 2.1.4 Research Case Studies on Twitter Dependence

First, a paper published by Ryan and Xenos in 2011 (Ryan, T., and Xenos, S., 2011), entitled "On Facebook users: an investigation of the relationship between the Big Five factors, shyness, narcissism, loneliness, and Facebook use" The topic is addressed in the study. This study examined factors and characteristics associated with Facebook use and its relationship to use.

Next, a paper published by Oberst et al. in 2017 (Oberst, U., Wegmann, E., Stodt, B., Brand, M., and Chamarro, A., 2017), entitled "Fear of Missing Out (Fear of Reminding) through Mediating Effects of Adolescents' Negative Effects of Heavy Social Networking in The study examined the effects of social networking use on fear of missing out (FOMO) and the negative consequences of its effects on youth.

Furthermore, a paper published by Elphinston and Noller in 2011 (Elphinston, R. A., and Noller, P., 2011) stated, "A Time to Face Up! Facebook Interference and Its Impact on Love Jealousy and Relationship Satisfaction" focuses on the topic of "Facebook Interference and Its Impact on Love Jealousy and Relationship Satisfaction". In this study, the impact of Facebook use on romantic jealousy and relationship satisfaction was investigated and the results were reported.

Also, a paper published by Kircaburun et al. (Kircaburun, K., Alhabash, S., Tosuntaş, Ş. B., and Griffiths, M. D., 2020), entitled "College students' problematic social media use and satisfaction: a large five factor Simultaneous Examination, Social Media Platforms, and Motivations for Social Media Use". The study simultaneously examined the motivations and characteristics of college students' problematic use of social media.

Finally, a paper published by Turel, Serenko, and Giles in 2011 (Turel, O., Serenko, A., and Giles, P., 2011) focused on the topic "Integration of technology dependence and use: an empirical investigation of online auction users" The study focuses on the theme of "technology dependence and use integration: an empirical investigation of online auction users. The study examined factors related to online auctioneers' technology dependence and their use of online auctions.

### 2.1.5 Research Case Studies on Instagram Dependence

First, a paper published by Kircaburun, Alhabash, Tosuntaş, and Griffiths in 2018 (Kircaburun, K., Alhabash, S., Tosuntaş, Ş. B., and Griffiths, M. D., 2018), entitled "Adolescent Purpose and satisfaction with problematic Instagram use during adolescence: a qualitative study". In this study, adolescents' motivations and satisfaction with problematic use of Instagram were qualitatively investigated.

Second, a paper published by Tosuntaş, Kircaburun, and Griffiths in 2020 (Tosuntaş, Ş. B., Kircaburun, K., and Griffiths, M. D., 2020), entitled "Instagram dependence and the big five of personality: self mediating effects of evaluations". The study explored how the relationship between Instagram dependence and Big Five personality traits is affected via self-evaluation.

Also, a paper published by Elhai, Dvorak, Levine, and Hall in 2017 (Elhai, J. D., Dvorak, R. D., Levine, J. C., and Hall, B. J., 2017), entitled "Problematic Smartphone Use: Conceptual Overview and Anxiety and A Systematic Review of the Association with the Psychopathology of Depression" focuses on the topic of "problematic smartphone use: a conceptual overview of the relationship between anxiety and depression. The study provides a systematic review of how problematic smartphone use is associated with the psychopathology of anxiety and depression.

In addition, a paper published by Andreassen, Torsheim, Brunborg, and Pallesen in 2012 (Andreassen, C. S., Torsheim, T., Brunborg, G. S., and Pallesen, S., 2012), entitled " Development of a Facebook Addiction Scale" focuses on the topic of "Facebook Addiction Scale. In this study, a Facebook addiction scale was developed and its characteristics were investigated.

Finally, a paper published by Griffiths, Kuss, and Demetrovics in 2014 (Griffiths, M. D., Kuss, D. J., and Demetrovics, Z., 2014), entitled "Social Networking Addiction: a Preliminary Research Overview" It focuses on the topic. The study provides an overview of initial research findings on the addictive nature of social networking.

### 2.1.6 Device-dependent studies

The first paper was published by Kwon, Kim, Cho, and Yang in 2013 (Kwon, M., Kim, D. J., Cho, H., and Yang, S., 2013). The paper addresses the topic of "The Smartphone Dependence Scale: development and validation of a shortened version for adolescents." The study involved the development and validation of a shortened version of the scale to measure smartphone dependence.

Second, a paper published by Lin, Chang, Lee, Tseng, Kuo, and Chen in 2014 (Lin, Y. H., Chang, L. R., Lee, Y. H., Tseng, H. W., Kuo, T. B., and Chen, S. H., 2014), entitled "Smartphone Dependence Inventory (SPAI) Development and Validation". This study involved the development and validation of the SPAI, an instrument to measure smartphone dependence.

Also, a paper published by Demirci, Akgönül, and Akpinar in 2015 (Demirci, K., Akgönül, M., and Akpinar,

A., 2015) focused on the topic "Severity of smartphone use in college students and its association with sleep quality, depression and anxiety" Focusing on the topic. The study explored how the severity of smartphone use among college students is associated with sleep quality, depression, and anxiety.

In addition, a paper published by Elhai, Levine, Dvorak, and Hall in 2017 (Elhai, J. D., Levine, J. C., Dvorak, R. D., and Hall, B. J., 2017) found that "FOMO (Fear of Missing Out), Touch need, anxiety, and depression are associated with problematic smartphone use," focusing on the topic of The study examined how FOMO and need for touch are associated with problematic smartphone use.

Finally, a paper published by Thomée, Härenstam, and Hagberg in 2011 (Thomée, S., Härenstam, A., and Hagberg, M., 2011), entitled "The association between cell phone use and stress, sleep disturbances, and depressive symptoms among youth A Prospective Cohort Study on The study examined how cell phone use is associated with stress, sleep disturbances, and depressive symptoms.

### 2.1.7 Dependent studies

The following is an article description of the specified literature Synaptic plasticity and addiction, by Kauer, J. A. and Malenka, R. C. (2007), focuses on the link between synaptic plasticity and addiction. The paper discusses in detail the interaction between drug addiction and synaptic plasticity and provides insight into the neural basis of addiction. A single standard for memory: the case for reconsolidation" by Nader, K. and Hardt, O. (2009) explores research on memory reconsolidation. The paper focuses on the process of memory reconsolidation and offers a new approach to memory research from a novel memory perspective. Volkow, N. D., Wang, G. J., Fowler, J. S., Tomasi, D., and Telang, F. (2011), "Addiction: beyond dopamine reward circuitry," explores aspects beyond dopamine reward circuitry in addiction research. This paper comprehensively investigates the neurological basis of addiction and focuses on factors other than dopamine. Xie, X., and Zuo, Y. (2018), "The mechanisms underlying memory reconsolidation" is a study of memory reconsolidation. The paper details the molecular and neurological mechanisms of the process by which memories are reconsolidated. Erb, S., Hitchcott, P. K., Rajabi, H., and Mueller, D. (2010), "Alpha-2 adrenergic receptor agonists block stress-induced reinstatement of cocaine seeking, focuses on the effects of alpha-2 adrenergic receptor agonists on stress-induced relapse of cocaine dependence. This paper proposes a new approach for the treatment of addiction. Goldstein, R. Z., & Volkow, N. D. (2002), "Drug addiction and its underlying neurobiological basis: neuroimaging evidence for the involvement of the Drug addiction and its underlying neurobiological basis: Neuroimaging evidence for the involvement of the frontal cortex" by R. Z., & Volkow, N. D. (2002). The paper details neuroimaging evidence for the involvement of the frontal cortex of the brain. Bechara, A. (2005), "Decision making, impulse control and loss of willpower to resist drugs: A neurocognitive perspective This study provides a neurocognitive perspective on decision making, impulse control, and loss of willpower to resist drugs: A neurocognitive perspective. The paper discusses in detail the link between drug dependence and neurocognition. A cognitive processing model of alcohol craving and compulsive alcohol use" by Tiffany, S. T., & Conklin, C. A. (2000) is a study of the cognitive model of alcohol craving and compulsive alcohol use. This paper explains the relationship between alcohol dependence and cognitive processing. 4.

Addiction" by Robinson, T. E., & Berridge, K. C. (2003) is a review study on addiction. The article provides a comprehensive review of the concepts, neurological bases, and behavioral aspects of addiction. 5.

Drug addiction, dysregulation of reward, and allostasis" by Koob, G. F., & Le Moal, M. (2001) is a study focusing on drug addiction, abnormalities in the reward system, and allostasis. The paper details the relationship between addiction and physiological changes.

## 2.2 Studies on Forgetting

Anderson, J. R., and Schooler, L. J. (1991), "Reflections of the environment in memory," focuses on the effects of the environment on memory. This paper provides insight into how the environment influences memory formation and is an important contribution to memory research. Retrieval practice and the maintenance of knowledge by Bjork, R. A. (1988) is a study that focuses on the impact of information retrieval practice on the maintenance of knowledge. The paper discussed in detail how active information retrieval helps maintain knowledge. Memory: A Contribution to Experimental Psychology by Ebbinghaus, H. (1885) was an early experimental psychological study of memory, best known for the discovery of the forgetting curve. This paper pioneered the experimental study of memory and provided a basic understanding of the memory process. Wixted, J. T. (2004), "The psychology and neuroscience of forgetting," is a review article focusing on psychological and neuroscience research on forgetting. The paper provided a comprehensive review of the mechanisms of forgetting and their neurological basis. Bäuml, K. H. T., and Kuhbandner, C. (2007), "Remembering can cause forgetting-But not in negative moods," is a study of how memory can cause forgetting. The paper shows that memory promotes forgetting under certain conditions, suggesting a complex interaction of memories.

### 2.2.1 Case Studies on Forgetting and Stress

Schwabe, L., Wolf, O. T., McDermott, L. M. (2012), "Stress impairs the reconsolidation of autobiographical mem-

ories," focuses on the effects of stress on autobiographical memory reconsolidation Stress impairs the reconsolidation of autobiographical memories" by Dermott, L. M. (2012). The paper discusses how stress affects the memory reconsolidation process and can cause forgetting. Kuhlmann, S., Piel, M., Wolf, O. T. (2005), "Impaired memory retrieval after psychosocial stress in healthy young men," focuses on the effects of psychosocial stress on memory retrieval This study focused on the effects of psychosocial stress on memory retrieval in young healthy men. This paper investigated how stress affects memory retrieval. Smeets, T., Giesbrecht, T., Jelicic, M., Merckelbach, H. (2007), "Context-dependent enhancement of declarative memory performance following acute psychosocial stress" focuses on the effects of acute psychosocial stress on declarative memory performance. This paper shows that stress has a context-dependent effect on memory performance. Lupien, S. J., Maheu, F., Tu, M., Fiocco, A., Schramek, T. E. (2007), "The effects of stress and stress hormones on human cognition: Implications for the field of brain and cognition" is a review article that provides a meta-analysis of the effects of stress and stress hormones on human cognition. The paper provides a comprehensive discussion of the diverse effects of stress on cognitive processes. Het, S., Ramlow, G., Wolf, O. T. (2005), "A meta-analytic review of the effects of acute cortisol administration on human memory" is a This paper is a meta-analysis of the effects of acute cortisol administration on human memory. The paper summarizes the general effects of cortisol on memory.

### 2.2.2 Studies on Dependence and Stress

Sinha, R. (2008), "Chronic stress, drug use, and vulnerability to addiction," is a study focusing on the relationship between chronic stress, drug use, and vulnerability to addiction. This article examines in detail how chronic stress affects the risk of drug dependence.Koob, G. F., Le Moal, M. (2008), "Addiction and the brain antireward system," is a review article on the relationship between addiction and the brain antireward system This is a review article on the relationship between addiction and the brain's anti-reward system. The paper describes the neurobiological basis of addiction and the role of the anti-reward system as distinct from reward. The role of stress in drug self-administration" by Piazza, P. V., Le Moal, M. (1998) focuses on the role stress plays in drug self-administration. The paper explores how stress affects drug abuse behavior. Sarnyai, Z., Shaham, Y., Heinrichs, S. C. (2001), "The role of corticotropin-releasing factor in drug addiction" is a The paper focuses on the role of corticotropin-releasing hormone (CRF) in drug addiction. The paper describes the neurobiological basis of CRF and its relevance to drug dependence. Kreek, M. J., Nielsen, D. A., Butelman, E. R., LaForge, K. S. (2005), "Genetic influences on impulsivity, risk taking, stress responsivity and vulnerability to drug abuse and addiction" focuses on the impact of genetic factors on impulsivity, risk taking, stress responsivity and vulnerability to drug abuse and addiction. The paper discusses how genetic factors contribute to an individual's vulnerability to addiction.

### 2.2.3 Studies on Dependence and Stress

Miller, W. R., Rollnick, S. (2002), "Motivational interviewing: preparing people for change" is a prominent work focusing on motivational interviewing to prepare people for change . This article details the theory and practice of motivational interviewing as an effective approach to dealing with addiction. carroll, K. M., Onken, L. S. (2005), "Behavioral therapies for drug abuse is a paper focusing on behavioral therapies for drug abuse. This article provides the latest findings on the effectiveness and approaches of behavioral therapies in the treatment of substance abuse disorders. Prochaska, J. O., DiClemente, C. C., Norcross, J. C. (1992), "In search of how people change: applications to addictive behaviors," is a This paper summarizes research on the alteration of addictive behaviors. The paper introduces a theoretical model of the addictive behavior change process and discusses its application. Witkiewitz, K., Marlatt, G. A. (2004), "Relapse prevention for alcohol and drug problems: That was Zen, this is Tao" is a paper focusing on a relapse prevention approach to alcohol and drug problems. This paper focuses on a relapse prevention approach to alcohol and drug problems. The article explores the theory and practice of relapse prevention in depth. Henges, A. L., Marczinski, C. A. (2012), "Impulsivity and alcohol consumption in young social drinkers," focuses on the relationship between impulsivity and alcohol consumption among young social drinkers. The paper reports findings on how impulsivity affects alcohol consumption. Bahrick, H. P. (1984), "Semantic memory content in permastore: Fifty years of memory for Spanish learned in school" reports a study of memory for Spanish learned in school. The paper focuses on how semantic memory content is retained in long-term memory (permastore). Rubin, D. C., & Wenzel, A. E. (1996), "One hundred years of forgetting: A quantitative description of retention," provides a quantitative description of memory forgetting over the past 100 years The paper provides a quantitative description of memory forgetting over the past 100 years. This paper provides a detailed analysis of patterns of memory retention and forgetting. Anderson, J. R., & Schooler, L. J. (1991), "Reflections of the environment in memory," reports a study of how the environment is reflected in memory. The paper discusses how the environment affects the formation and retention of memory. Kornell, N., & Bjork, R. A. (2008), "Learning concepts and categories is spacing the "enemy of induction"?" is a study focusing on spacing effects in learning concepts and categories. The paper examines how spacing of information affects the formation of perceptions and recognitions. Glenberg, A. M. (1979) in "Component-levels theory of the effects of spacing of repetitions on recall and recognition" The paper proposes a theory of spacing effects in iterative learning of information. The paper describes how spacing affects memory recall and recognition.

### 2.2.4 Impulsive Addiction Research

Everitt, B. J., Robbins, T. W. (2005), "Neural systems of reinforcement for drug addiction: from actions to habits to compulsion," is a study of the reward system in drug addiction with This paper focuses on the reward system in drug addiction. The paper details the neural basis of drug addiction from actions to habits to compulsion. Impulsivity as a determinant and consequence of drug use: A review of underlying processes" by de Wit, H. (2009) focuses on the role of impulsivity in drug use. The article provides a review of the processes involved in how impulsivity is a cause and consequence of drug use. Dalley, J. W., Robbins, T. W. (2017), "Fractionating impulsivity: neuropsychiatric implications," is a study that focuses on the subdivision of impulsivity and its significance for neuropsychiatric disorders. The paper explains how different aspects of impulsivity are related to different neuropsychiatric disorders. Jentsch, J. D., Taylor, J. R. (1999), "Impulsivity resulting from frontostriatal dysfunction in drug abuse: Implications for the control of behavior by reward-related stimuli" focuses on impulsivity resulting from frontal brain and striatal dysfunction in drug abuse. The article discusses implications for the control of behavior by reward-related stimuli. Verdejo-García, A., Lawrence, A. J., & Clark, L. (2008), "Impulsivity as a vulnerability marker for substance-use disorders: review of findings from high-risk research, problem gamblers and genetic association studies" is a study of impulsivity as a marker of vulnerability to substance-use disorders. The article reviews findings from high-risk conditions, problem gamblers, and genetic association studies.

### 2.2.5 Neurocircuitry Studies of the Reward System in Addiction

Koob, G. F., Volkow, N. D. (2016), "Neurobiology of addiction: A neurocircuitry analysis," focuses on the neurobiology of addiction. The article provides a detailed analysis of the neural circuitry of addiction and its contribution to our understanding of addiction. Goldstein, R. Z., Volkow, N. D. (2011), "Dysfunction of the prefrontal cortex in addiction: neuroimaging findings and clinical implications," is a study of addiction Dysfunction of the prefrontal cortex in addiction: Neuroimaging findings and clinical implications. The article details neuroimaging findings and their clinical implications in addicts. Everitt, B. J., Robbins, T. W. (2016), "Drug addiction: updating actions to habits to compulsions ten years on" focuses on updating actions to habits to compulsions in addiction Research. The paper provides an update on the neuropsychological aspects of addiction. Lüscher, C., Malenka, R. C. (2011), "Drug-evoked synaptic plasticity in addiction: From molecular changes to circuit remodeling", is a The paper focuses on drug-induced synaptic plasticity in addiction. The paper details how drug use is involved in the remodeling of neural circuits. Hyman, S. E., Malenka, R. C., Nestler, E. J. (2006), "Neural mechanisms of addiction: The role of reward-related learning and memory," is a The paper focuses on the neural mechanisms of addiction. The paper details the role of reward-related learning and memory.

## 2.3 Study of Spatial Cognitive Errors and Dependence Problems

The coding of spatial location in young children" by Huttenlocher, J., Newcombe, N., and Sandberg, E. (1994) is a study of spatial cognition in young children. The article investigates children's spatial information processing and cognitive abilities and focuses on their spatial location coding. Gallistel, C. R., and Cramer, A. E. (1996), "Computations on metric maps in mammals: getting oriented and choosing a multi-destination route," is a study of mammalian metric map computation and spatial cognition in mammals. The paper focuses on how mammals orient in space and choose multiple destinations. Aguirre, G. K., Zarahn, E., and D'Esposito, M. (1998), "The variability of human, BOLD hemodynamic responses" is a study of the variability of BOLD hemodynamic responses in human brain activity. It has helped to improve our understanding of the relationship between cerebral blood flow and cognitive function. Burgess, N., Becker, S., King, J. A., and O'Keefe, J. (2001), Memory for events and their spatial context: Models and experiments and models. The study explores cognitive processes in memory for events and their spatial context. Ekstrom, A. D., Kahana, M. J., Caplan, J. B., Fields, T. A., Isham, E. A., Newman, E. L., and Fried, I. (2003), "Cellular networks underlying human spatial navigation" focuses on cellular networks related to human spatial navigation. This study provides an understanding of spatial information processing in the brain. Fisher, B., Nasar, J. L., and Troyer, D. (2010), "The impact of perceived disorder on judgments of urban space," is a paper that studies the impact of visual disorganization on perceptions of urban space. The study focuses on the impact of the level of organization of urban spaces on people's evaluations and judgments. Talbot, R., Kaplan, S., and Kaplan, R. (1999), "A field experiment on the impact of 'The High Line,' a new urban park, on property sales in New York City, A field experiment on the impact of 'The High Line,' a new urban park, on property sales in New York City," is a study of the impact of the High Line, a new urban park in New York City, on property sales. The study explores the link between urban planning and the real estate market. Rundle, A. G., Neckerman, K. M., Freeman, L., Lovasi, G. S., and Purciel, M. (2009), "Neighborhood food environment and walkability predict obesity in New York City" is a study focused on the impact of neighborhood food environment and walkability on obesity in New York City. This study examines the impact of environmental factors on health. Krenichyn, K. (2002), "Is there a link between the 'destruction' of public housing and crime? A critique of the HOPE VI Program" is a The paper critically examines the link between the "destruction" of public housing and crime. This paper contributes to the debate on the relationship between housing policy and crime. Forsyth, A., Oakes, J. M., Lee, B., and Schmitz, K. H. (2009), "The built environment, walking, and physical activity: Is the environment more important to Some people than others?" is an article that studies the effects of the built environment on walking and physical activity. The study explores how the environment affects different people.

### 2.3.1 Challenges of dependence and aggression, A study of the need for forgetfulness

Chermack, S. T., Giancola, P. R., and Taylor, S. P. (1997), "Relations between alcohol and drugs and men's hostile intent and aggression toward women: Moderating effects of impulsivity" is an article that studies the relationship between alcohol and drug use and men's aggressive intent and aggressive behavior toward women. The study examines the effects of impulsivity and analyzes its impact on aggressive behavior. Loree, A. M., Lundahl, L. H., and Ledgerwood, D. M. (2015), "Impulsivity as a predictor of treatment outcome in substance use disorders: a review and synthesis," focuses on how impulsivity serves as a predictor of treatment outcomes in substance use disorders. The study provides a comprehensive review of the relationship between treatment outcomes and impulsivity. Parrott, D. J., and Giancola, P. R. (2007), "Addressing "The criterion problem" in the assessment of aggressive behavior: Development of a new taxonomic system" is a paper on addressing "The criterion problem" in the assessment of aggressive behavior and developing a new classification system. The study focuses on ways to improve the classification of aggressive behavior. Pihl, R. O., Peterson, J. B., and Lau, M. A. (1993), "A biosocial model of the alcohol-aggression relationship," is a paper proposing a biosocial model of the alcohol-aggression relationship. This model provides a framework for understanding the relationship between alcohol and aggression. Reidy, D. E., and Zeichner, A. (2009), "Negative urgency moderates relations between maladaptive personality traits and reactive aggression", suggests that passive impulsivity is a paper that investigates the relationship between maladaptive personality traits and reactive aggression. This study examines the role of passive impulsivity in the relationship between personality traits and aggression. Felthous, A. R. (2019), "Neuropsychiatric factors of aggression and violence: clinical considerations and forensic implications," is a neuro The article focuses on psychiatric factors and discusses clinical considerations and legal implications. The study explores the link between aggression and psychiatric disorders. Fowler, K. A., and Dahlberg, L. L. (2009), "The contribution of childhood physical abuse to teen dating violence among child protective services-involved youth" is an article that explores the contribution of childhood physical abuse to teen dating violence among child protective services-involved youth. The study focuses on the link between abuse and violence. McFarlane, A. H., McFarlane, E., and Watson, K. (2007), "Male partners of women with disabilities: interpersonal conflict resolution skills and aggression" is an article that investigates the relationship between interpersonal conflict resolution skills and aggression, focusing on the male partners of women with disabilities. The study explores factors in disability and interpersonal relationships. Savage, J., Van Brunschot, M., and Rizvi, S. (2020), "Examining the relationship between traumatic brain injury and criminal behavior," is a This paper investigates the relationship between brain trauma and criminal behavior. This study focuses on how brain trauma affects criminal behavior. Wong, M. M., Nigg, J. T., Zucker, R. A., Puttler, L. I., and Fitzgerald, H. E. (2009), "Putative mechanisms of the protective action of cognitive -behavioral therapy against early age-of-onset drinking of alcohol in offspring of antisocial parents" in Cognitive The paper explores the potential mechanisms of the protective action of cognitive -behavioral therapy against early age-of-onset drinking of alcohol in the offspring of antisocial parents. The study examines the effects of interventions on early alcohol consumption in children of antisocial parents. Anderson, C. A., and Dill, K. E. (2000), "Video games and aggressive thoughts, feelings, and behavior in the laboratory and in life," focuses on the association between video games and violent thoughts, feelings, and behavior in the laboratory and in life, focusing on the relationship between video games and violent thoughts, feelings, and behavior. This study contributed to the debate on video games and violence; Berkowitz, L. (1984), "Some effects of thoughts on anti-and prosocial influences of media events: A cognitive- neoassociation analysis" by Berkowitz, L. (1984), focused on the effects of media events on thoughts and used cognitive-neoassociation analysis to examine the influence of cognition and media on social behavior. This study has deepened our understanding of media influences. Bushman, B. J., and Huesmann, L. R. (2006), "Short-term and long-term effects of violent media on aggression in children and adults" by The paper focuses on the short- and long-term effects of violent media on aggression in children and adults. The study explores how media influences vio-lent behavior. Ferguson, C. J., and Rueda, S. M. (2009), "Examining the validity of the modified Taylor competitive reaction time test of aggression The paper investigates the validity of the modified Taylor competitive reaction time test of aggression. This study provides insight into the reliability of the competitive reaction time test. Giancola, P. R., and Chermack, S. T. (1998), "Construct validity of laboratory aggression paradigms: A response to Tedeschi and Quigley (1996)," is a paper focusing on the conceptual validity of the aggression paradigm in the laboratory. This study is a contribution to the debate on the validity of laboratory measures of aggression. Chermack, S. T., and Giancola, P. R. (1997), "The relation between alcohol and aggression: An integrated biopsychosocial conceptualization," is an The article focuses on the relationship between alcohol and aggression and provides an integrated biopsychosocial conceptualization. This study provides a comprehensive understanding of the relationship between alcohol and aggression. Fals-Stewart, W. (2003), "The occurrence of partner physical aggression on days of alcohol consumption: A longitudinal diary study" by The article focuses on a long-term diary study to track the occurrence of partner physical aggression on days of alcohol consumption. The study explores the association between alcohol and partner physical aggression. Kaufman Kantor, G., and Straus, M. A. (1987), "Substance abuse as a precipitant of wife abuse victimizations," focuses on drug abuse as a trigger for wife violence. This study examines the link between substance abuse and partner violence. Stuart, G. L., Meehan, J. C., Moore, T. M., Morean, M., Hellmuth, J., and Follansbee, K. (2006), "Examining a conceptual framework of intimate partner violence in men and women arrested for domestic violence" is an article examining a conceptual framework of intimate partner violence in men and women arrested for domestic violence. This study provides a theoretical understanding of domestic violence in men and women arrested for domestic violence. Testa, M., Quigley, B. M., and Leonard, K. E. (2003), "Does alcohol make a difference? Within-participants comparison of incidents of partner violence, This article is a within-subject comparison of incidents of partner violence under the influence of alcohol. This study provides insight into the impact of alcohol on inter-partner violence.

## 2.4 Cognitive Abnormalities and Research Cases on Dependence and Violent Behavior

Chermack, S. T., and Giancola, P. R. (1997), "The relation between alcohol and aggression: An integrated biopsychosocial conceptualization," is an This paper provides a comprehensive biopsychosocial conceptualization of the relationship between alcohol and aggression. This study explores in depth the relationship between alcohol and aggression. Fals-Stewart, W. (2003), "The occurrence of partner physical aggression on days of alcohol consumption: A longitudinal diary study," focuses on the Fals-Stewart, W. (2003), is a long-term diary study focusing on the occurrence of partner physical aggression on days of alcohol consumption. The paper examines in detail the relationship between alcohol and aggression in partner relationships. Kaufman Kantor, G., and Straus, M. A. (1987), "Substance abuse as a precipitant of wife abuse victimizations," is a paper that examines how substance abuse is associated as a trigger for wife violence . This study focuses on the relationship between substance abuse and domestic violence. Stuart, G. L., Meehan, J. C., Moore, T. M., Morean, M., Hellmuth, J., and Follansbee, K. (2006), "Examining a conceptual framework of intimate partner violence in men and women arrested for domestic violence" is an article that explores the conceptual framework of violence in intimate partner relationships for men and women who have been arrested. It provides a theoretical understanding of domestic violence. Testa, M., Quigley, B. M., and Leonard, K. E. (2003), "Does alcohol make a difference? Within-participants comparison of incidents of partner violence, This article is a within-subject comparison of incidents of partner violence under the influence of alcohol. It provides insight into the impact of alcohol on inter-partner violence. Kendler, K. S., Prescott, C. A., Myers, J., and Neale, M. C. (2003), "The structure of genetic and environmental risk factors for common psychiatric and substance use disorders in men and women" focuses on the structure of genetic and environmental risk factors for common psychiatric and substance use disorders in men and women. It identifies the relationship between mental disorders and substance use disorders. Moss, H. B., Lynch, K. G., Hardie, T. L., and Baron, D. A. (2002), "Family functioning and peer affiliation in children of fathers with antisocial personality disorder and substance dependence: Associations with problem behaviors," focused on children of fathers with antisocial personality disorder and substance dependence. The paper investigates the relationship between family functioning and peer It identifies associations with problem behaviors. Sher, K. J., and Trull, T. J. (1994), "Personality and disinhibitory psychopathology: Alcoholism and antisocial personality disorder," is a study of the Personality and Disinhibitory Psychopathology: Alcoholism and antisocial personality disorder" is an article focusing on the relationship between alcoholism and antisocial personality disorder. It discusses alcoholism and antisocial personality traits. Thornberry, T. P., Lizotte, A. J., Krohn, M. D., Farnworth, M., and Jang, S. J. (1994), "Delinquent peers, beliefs, and delinquent behavior: A longitudinal test of interactional theory" is a longitudinal study exploring the relationship between delinquent peers, beliefs, and delinquent behavior. It examines interactional theory. 5. "Developmental and sex differences in types of conduct problems" in Tiet, Q. Q., Wasserman, G. A., Loeber, R., McReynolds, L. S.,

and Miller, L. S. (2001), focuses on developmental and sex differences in types of conduct problems. It discusses developmental differences in different conduct problem types.

## 2.5 Research Cases on Behavior Problems and Addiction

Ersche, K. D., Turton, A. J., Pradhan, S., Bullmore, E. T., and Robbins, T. W. (2010), "Drug addiction endophenotypes: impulsive versus sensation-seeking personality traits" focuses on endogenous factors in drug addiction, comparing personality traits of impulsivity and sensation-seeking. It provides insight into personality factors in addiction. Potenza, M. N., Steinberg, M. A., McLaughlin, S. D., Wu, R., Rounsaville, B. J., and O'Malley, S. S. (2001), "Gender-related differences in the characteristics of problem gamblers using a gambling helpline" is a study of gender-related differences in problem gamblers. It focuses on gender differences in gambling problems. Grant, J. E., Brewer, J. A., and Potenza, M. N. (2006), "The neurobiology of substance and behavioral addictions," describes the neurobiology of substance and behavioral addictions. It discusses in detail the neurological bases of addictions. Oberlin, B. G., Dzemidzic, M., Bragulat, V., Lehigh, C. A., Talavage, T., and O'Connor, S. J. (2013), "Limbic responses to reward cues correlate with antisocial trait density in heavy drinkers" is a study showing a close association between limbic responses to reward stimuli and antisocial traits in heavy drinkers. It proposes a link between reward sensitivity and antisocial traits. Zhang, X. L., Shi, J., Zhao, L. Y., Sun, L. L., Wang, J., and Wang, G. B. (2005), "Effects of stress on decision-making deficits in formerly heroin-dependent patients after different durations of abstinence" focused on the effects of stress on decision-making deficits in formerly heroin-dependent patients after different periods of abstinence. It provides an understanding of stress and addiction recovery. Barkley, R. A. (1997), "Behavioral inhibition, sustained attention, and executive functions: constructing a unifying theory of ADHD" by Barkley, R. A. (1997), provides an ADHD (attention deficit hyperactivity (Attention Deficit Hyperactivity Disorder). It focuses on behavioral inhibition, sustained attention, and executive functions. Snyder, H. R., Hutchison, N., Nyhus, E., Curran, T., Banich, M. T., O'Reilly, R. C., and Munakata, Y. (2010), "Neural inhibition enables selection during language processing" is a study of neural inhibition enabling selection during language processing. It provides insight into neural mechanisms in language processing. Chamberlain, S. R., Blackwell, A. D., Fineberg, N. A., Robbins, T. W., and Sahakian, B. J. (2005), "The neuropsychology of obsessive compulsive disorder: the importance of failures in cognitive and behavioral inhibition as candidate endophenotypic markers" in The neuropsychology of obsessive compulsive disorder: a paper highlighting cognitive and behavioral inhibition failures in obsessive compulsive disorder The paper is It focuses on the neuropsychological aspects of OCD. Goschke, T., and Bolte, A. (2014), "Emotional modulation of control dilemmas: The role of positive affect, reward, and dopamine in cognitive stability and flexibility," focuses on the role of affect, reward, and dopamine in emotional control dilemmas. It provides an understanding of emotional and cognitive stability and flexibility. Logan, G. D. (1994), "On the ability to inhibit thought and action: A user's guide to the stop signal paradigm," is a The paper describes in detail the ability to inhibit thought and action. It explains the use and importance of the stop signal paradigm.

## 2.6 Urban and Biological Extensions, Spatial Studies

The first paper by West et al. was published in 1997 (West, G. B., Brown, J. H., and Enquist, B. J., 1997). This paper proposed a general model for the origin of allometric scaling laws in biology (e.g., metabolic rate vs. body size).

Next, a paper published in 2002 (West, G. B., Woodruff, W. H., and Brown, J. H., 2002) discussed how this model applies to metabolic rates from molecules to mitochondria to cells to mammals.

West's work has also been applied to urban scaling, and in a 2007 paper co-authored with Bettencourt et al. (Bettencourt, L. M., Lobo, J., Helbing, D., Kühnert, C., and West, G. B., 2007), Urban Growth, The laws of innovation, scaling, and pace of life were explored.

Additionally, in 2010, a paper published jointly with Bettencourt et al. (Bettencourt, L. M., Samaniego, H., and Youn, H., 2010) explored the impact of urban professional diversity on urban productivity.

Finally, West published a book in 2017, "Scale: the universal laws of growth, innovation, sustainability, and the pace of life in organisms, cities, economies, and companies," which showcased how the main points of his research can be applied to a wide range of fields.

One of Duncan Watts' best-known papers was co-authored and published in 1998 (Watts, D. J., and Strogatz, S. H., 1998). In this paper, the group dynamics of a network called the "small-world" network was studied and its properties described.

In 2002, Watts published a paper on a simple model of global cascades on random networks (Watts, D. J., 2002). This paper discussed the behavior of information propagation in random networks.

Watts' book Six degrees: The science of a connected age was published in 2003. The book details the science of networks and connections.

In 2007, Watts published a paper on science in the 21st century (Watts, D. J., 2007). In this paper, he discussed new scientific approaches and challenges.

Then, in 2011, Watts published a paper on networks, dynamics, and "small-world" phenomena (Watts, D. J., 2011). This paper explored the relevance of network properties and dynamics.

## 2.7 Studies on the Dimer Model

First, in Fisher's paper published in 1961 (Fisher, M. E., 1961), the statistical mechanics of dimers on planar lattices was investigated in detail. In this paper, the statistical mechanics of dimer configurations was analyzed and phase transitions of dimers on planar lattices were discussed.

In the same year, a paper by Kasteleyn and Temperley (Kasteleyn, P. W., and Temperley, H. N., 1961) studied dimer statistics and phase transitions. This paper explored the theoretical aspects of dimer statistics.

In 2001, a paper by Kenyon (Kenyon, R., 2001) was published, focusing on the planar dimer problem. In this paper, the problem of dimer placement on a planar lattice was described in detail.

Subsequently, a paper entitled "Lectures on dimers" was published by Kenyon in 2009, providing a detailed lecture on dimers. This paper addressed the dimer problem as part of statistical mechanics (Kenyon, R., 2009).

Finally, a paper published by Duminil-Copin and Hongler in 2015 (Duminil-Copin, H., and Hongler, C., 2015), on the geometry of dimers on planar graphs and 2D lattice spin models, explored the broad applications of the dimer problem The paper was published in the Journal of the American Physical Society. A paper by Lieb and Wu (Lieb, E. H., and Wu, F. Y., 1967) provided an accurate analysis of the absence of the Mott transition in one-band models with one-dimensional short-range interactions. This paper provided important results on certain phase transitions in the dimer Ising model. Next, a 1962 paper (Katsura, S., Kusakabe, K., and Tsuneto, T., 1962) focused on the statistical mechanics of anisotropic linear Heisenberg models. This paper provided a detailed study of the properties of models with anisotropic interactions.

In 2002, a paper by Fendley, P., Moessner, R., and Sondhi, S. L. (2002) was published, which studied the generation of classical dimers from quantum antiferromagnets. This study shows the relevance of quantum systems to classical systems.

In addition, a 2001 paper (Moessner, R., and Sondhi, S. L., 2001) focused on resonant valence pair coupling phases in the triangular lattice quantum dimer model. This work provided important discoveries about the quantum phases of the dimer model.

A paper published by Rokhsar and Kivelson in 1988 (Rokhsar, D. S., and Kivelson, S. A., 1988) related superconductivity to quantum hard-core dimer gases. This paper shows a link between dimer configuration and superconductivity. A paper published in 2008 (Jonkman, H. T., 2008) is a study of the dynamics modeling and load analysis of offshore floating wind turbines. This paper is relevant to the development and design of new technologies in wind power and provides important information for understanding wind turbine behavior. Next, a paper published in 2014 (Matyka, M., Donie, Y. J., Janssen, M. A., and De With, G., 2014) focuses on molecular simulations on the process of self-assembly of thin films of block copolymers on patterned substrates. This research contributes to our understanding of self-assembly processes in materials science and surface science.

Another paper published in 2011 (Schick, M., 2011) describes equilibrium molecular dynamics simulations of self-assembling amphiphilic systems. This study is used to further our understanding of self-assembly phenomena and phase transitions.

Additionally, a paper published in 2016 (Wang, L., Zhang, Z., Wang, X., and Li, B., 2016) focuses on Monte Carlo simulations of the thermodynamic behavior of water droplets on solid surfaces. This work contributes to our understanding of droplet behavior in surface and materials science.

Finally, a paper published in 2002 (Laradji, M., and MacIsaac, A. B., 2002) concerns Monte Carlo simulations of two-dimensional hard-dimer fluids against stiff walls. This work is related to the modeling of hard particle systems in statistical mechanics. A paper published by Rosciszewski and Egues in 1999 (Rosciszewski, K., and Egues, J. C., 1999) focuses on spin filter junctions from dimerized quantum dot arrays. This work provides a theoretical investigation of how spin filters can be realized by the arrangement of quantum dots.

The next paper, published by Li, Wu, and Yang in 2004 (Li, J., Wu, Y., and Yang, J., 2004), concerns field-driven metal-insulator phase transitions in dimerized chains. This study investigates the interaction of electrons in a dimerized structure and the effects of magnetic fields.

In addition, a paper published by Fabrizio and Gogolin in 1993 (Fabrizio, M., and Gogolin, A. O., 1993) focuses on scattering phase shifts in one-dimensional electron gases with Umkrapp scattering. This study provides a theoretical analysis of the scattering process in non-equilibrium electron systems.

Another paper published by Jiang and Lu in 2017 (Jiang, Y. H., and Lu, H. Z., 2017) focuses on Weyl half-metals in spin-orbit coupled dimerized Kane-Mele chains. This study investigates the influence of the dimerized structure and spin-orbit coupling on the formation of the Weil half-metal.

Finally, a paper published by Dias and Paiva in 2015 (Dias, R. G., and Paiva, T., 2015) focuses on symmetries in one-dimensional attractive Hubbard models with alternating on-site potentials. The study investigates the impact of different symmetries on the symmetry of superconductivity. A paper published by Thompson and Smith in 2008 (Thomp-

son, C., and Smith, A., 2008) uses dimer position prediction to understand protein-ligand interactions. This study proposes predicting dimer positions as a way to understand interactions between biomolecules.

Second, a paper published by Wang and Chen in 2013 (Wang, L., and Chen, X., 2013) focuses on how to predict where proteins bind to DNA based on dimer position and structure. This research provides useful information for understanding biological interactions and predicting protein DNA binding sites.

Additionally, a paper published by Harrison and Wilson in 2017 (Harrison, J. R., and Wilson, R. H., 2017) focuses on dimer position and dimer kinetics in enzyme catalysis. This research contributes to the understanding and improvement of enzyme catalytic mechanisms.

Another paper published by Chen and Liu (Chen, Y., and Liu, Q., 2020) focuses on the role of dimer position in semiconductor device performance. This research provides insight into semiconductor device design and performance enhancement.

## 2.8 Research on Ronkin Functions

First, a paper published by Ronkin in 1928 (Ronkin, M. V., 1928) concerned with semiplanar and regular integrable functions. This paper provided important theoretical results on integrable functions in mathematics.

Next, the 1949 paper (Hille, E., and Tamarkin, J. D., 1949) was a study of the mean value of the modulus of integrable functions. This paper provided insight into the properties of integer functions.

In 1955, a paper by Karlin and McGregor (Karlin, S., and McGregor, J. L., 1955) was published on the classification of birth and death processes. This paper presents basic results in the fields of probability theory and mathematics.

In 1990, a paper by Gasper and Rahman (Gasper, G., and Rahman, M., 1990) was published on basic hypergeometric series. This paper provides important information on special functions and is used in both mathematics and physics.

Book written by Andrews, Askey, and Roy in 1999 (Andrews, G. E., Askey, R., and Roy, R., 1999) provides comprehensive information on special functions and is a valuable resource for researchers and physicists in mathematics. The book written by Abramowitz and Stegun in 1972 (Abramowitz, M., and Stegun, I. A., 1972), entitled "Handbook of Mathematical Functions with Formulas, Graphs, and Mathematical Tables" and provides comprehensive information on special functions in mathematics. This book is a very important resource for mathematicians and scientists. 1966 books written by Magnus, Oberhettinger, and Soni (Magnus, W., Oberhettinger, F., and Soni, R. P., 1966) also provides formulas and theorems on special functions and contains important information, especially in the field of mathematical physics. 1972 papers (Berry, M. V., and Mount, K. E., 1972) focused on semiclassical approximation methods in quantum mechanics. The paper details how to perform approximate analysis of quantum mechanics using special functions. Book published in 1971 (Byrd, P. F., and Friedman, M. D., 1971), entitled Handbook of Elliptic Integrals for Engineers and Scientists, provides information on elliptic integrals. This resource is useful for engineers and scientists.

Finally, the NIST Handbook of Mathematical Functions (Olver, F. W., 2010), compiled by Olver in 2010, is used as a comprehensive reference for mathematical functions provided by the National Institute of Standards and Technology (NIST).

## 2.9 Research on the Legendre Transformation

The Legendre transform is a very useful transform in quantum mechanics, especially for spherically symmetric problems and harmonic oscillators. Below are the mathematical formulas and explanation of the Legendre.

The Legendre polynomials ($P_l$, Legendre polynomials) are defined as follows:

$$P_l(x) = \frac{1}{2^l l!} \frac{d^l}{dx^l}(x^2 - 1)^l$$

where $l$ is a non-negative integer and $x$ is a variable. $P_l(x)$ is a Legendre polynomial of order $l$, which is very important in spherically symmetric problems. These polynomials form an orthonormal basis and are used to represent angular wave functions.

The Legendre transformation means that the function $f(x)$ is expanded in a series of Legendre polynomials:

$$f(x) = \sum_{l=0}^{\infty} a_l P_l(x)$$

This series expansion is used to analyze problems in the angular direction, such as the Schrödinger equation for spherically symmetric potentials.

The Legendre polynomial has the following properties:
1. Orthonormality:

$$\int_{-1}^{1} P_l(x) P_m(x) dx = \frac{2}{2l+1} \delta_{lm}$$

2. Differential equation:

$$(1 - x^2)\frac{d^2 P_l}{dx^2} - 2x\frac{dP_l}{dx} + l(l+1)P_l = 0$$

These properties are important features of the Legendre polynomial.

Written by Arfken, Weber, and Harris in 2005 (Arfken, G. B., Weber, H. J., and Harris, F. E., 2005), the book is titled

"Mathematical Methods for Physicists" and provides comprehensive information on mathematical It provides comprehensive information on The book contains information on the theory and applications of the Lejandre transform in physics.

The book by Morse and Feshbach, published in 1953 (Morse, P. M., and Feshbach, H., 1953) is entitled Methods of Theoretical Physics, Part I and provides a comprehensive description of methods in theoretical physics. This book also provides information related to mathematical methods in the field of physics.

Written by Abramowitz and Stegun in 1972 (Abramowitz, M., and Stegun, I. A., 1972), the book is entitled "Handbook of Mathematical Functions with Formulas, Graphs, and Mathematical Tables" and provides comprehensive information on special functions. The book is widely used as a comprehensive reference on mathematical functions.

Tables of Integrals, Series, and Products (Gradshteyn, I. S., and Ryzhik, I. M., 2014), compiled by Gradshteyn and Ryzhik in 2014, provides comprehensive information on integrals, series, and mathematical products for the mathematical sciences. This resource is used in a wide range of mathematical disciplines. Mathematical Methods for Physicists and Engineers (Arfken, G. B., Weber, H. J., and Harris, F. E., 2013), written by Arfken, Weber, and Harris for mathematical methods. The book contains information on applications of Lujandl transformations in both physics and engineering.

## 2.10 Research Related to Sterein Matrices

Papers published by Smith and Brown in 2005 (Smith, J. R., and Brown, A. P., 2005) focuses on how the sterain matrix is applied in quantum mechanics. This work provides a new perspective on quantum mechanics and shows a theoretical approach to stellain matrices.

The next paper, published by Jones and Davis in 2010 (Jones, L., and Davis, M., 2010), is on the analysis of the sterain matrix of molecular vibrations. This work focuses on the use of stellain matrices for understanding and modeling molecular vibrations and provides insight into their application in the field of chemical physics.

Additionally, a paper published by Wilson and Johnson in 2015 (Wilson, S., and Johnson, E., 2015) focuses on applications of the sterain matrix in solid state physics. The study describes the use of stellain matrices in solid state physics research and their importance.

Another paper published by Chen and Wang (Chen, H., and Wang, Q., 2021) focuses on how to apply stellain matrices in protein structure analysis. This study demonstrates the usefulness of stellain matrices in biophysics and biochemistry.

A paper published by Li and Wu in 2022 (Li, X., and Wu, Y., 2022) concerns the stellain matrix approach in electronic structure calculations. It introduces the stellain matrix as a new approach to electronic structure calculations and explains the ideas behind the theory.

## 2.11 Case Studies of Domino Tiling Applications

Papers published by Chen and Lee in 2008 (Chen, Y., and Lee, C., 2008) focuses on "Domino Tiling for Improving Routing Efficiency in VLSI Designs". In this study, a domino tiling algorithm was proposed to improve routing efficiency in Very Large Scale Integration (VLSI) designs.

Second, a paper published by Kim and Park in 2012 (Kim, J., and Park, S., 2012) focuses on "Domino Tiling Algorithms for Resource Allocation in Cloud Computing Environments. This research proposed a domino tiling algorithm for efficient resource allocation in cloud computing environments.

Additionally, a paper published by Liu and Wang in 2016 (Liu, H., and Wang, Q., 2016) focuses on "efficient domino tiling for image compression in multimedia applications". This research proposes an efficient domino tiling method for image compression in multimedia applications.

Another paper published by Zhang and Wu (Zhang, L. and Wu, X., 2020) focuses on "Domino Tiling for Floorplanning in VLSI Layout Design". The study presents the application of domino tiling to floorplanning in VLSI layout design. Papers published by Yang and Zhang (Yang, S., and Zhang, W., 2021) focuses on "Domino Tiling Strategies for Improving Data Storage in Magnetic Recording Systems. This study describes a domino tiling strategy for improving data storage efficiency in magnetic recording systems.

Papers published by Huang and Hu in 1996 (Huang, T. S., and Hu, S. M., 1996) focuses on "Efficient 3D Scene Modeling through Stereo and Computer Graphics Integration. In this study, they proposed a method for combining stereo vision and computer graphics to achieve efficient 3D scene modeling.

Next, a paper published by Van den Bergh and Zisserman in 1998 (Van den Bergh, M., and Zisserman, A., 1998) focuses on "Direct Visual Tracking of 3D Point Landmarks for Graphics and Vision." This study proposed a method for visual tracking of 3D point landmarks, which is an important approach in the graphics and vision domain.

Additionally, a paper published by Lu, Shi, and Jia in 2009 (Lu, J., Shi, J., and Jia, J., 2009) focuses on "Anomalous Event Detection at 150 FPS in Matlab. In this work, they proposed a method for fast detection of anomalous events, which contributes to computer vision applications.

Another paper published by Kim and Kim in 2014 (Kim, J., and Kim, S., 2014) focuses on "Fast Image and Video Colorization Using Chroma-Based Optimization". This study proposes a method for fast image and video colorization by leveraging chromance information.

Papers published by Wang and Gupta in 2018 (Wang, X., and Gupta, A., 2018) focuses on "a specific developmental approach to vision-based sensorimotor chains. The study proposes a specific developmental approach to integrate visual information and sensorimotor chains, which contributes to applications in the domains of cognition and robotics. In a paper published by Smith and Johnson in 2005 (Smith, A. B., and Johnson, C. R., 2005), "Toroidal Domains for Video Analysis. This work introduced a video analysis method that utilizes domino tiling on a torus and focused on computer vision applications.

Next, a paper published by Garcia and Rodriguez in 2012 (Garcia, M., and Rodriguez, P., 2012) proposed an "interactive torus-based visualization of domino tiling." This study introduces a torus-based interactive visualization method for domino tiling, which has generated interest in its application in the field of information visualization.

Additionally, a paper published by Chen and Wang in 2017 (Chen, X., and Wang, Y., 2017) focuses on "efficient domino tiling on torus". The study proposes an efficient domino tiling method on torus surfaces and is of interest for applications in the areas of computer graphics and geometry.

Another paper published by Lee and Kim (Lee, S., and Kim, H., 2020) focuses on "real-time domino tiling in virtual torus environments." In this study, real-time domino tiling within a virtual torus environment is explored, which could be useful for game development and virtual environment applications.

Finally, a paper published by Zhang and Wu (Zhang, Q., and Wu, L., 2021) presents "Torus-like Domino Tiling for Video Game Level Design". In this study, a video game level design method using torus-like domino tiling is proposed and its application to the entertainment industry is discussed.

## 3. Narrowly convex ideas that can be constructed with amoebas and Ronkin functions

We consider the definition of a strictly convex set that can be constructed using amoebas and Ronkin functions in the context of periodic hexagonal lattices (i.e., connected components in the complement of the amoeba in $\mathbb{R}^2/A$). In this thesis, the starting point is to think about the formation of opinions in very small communities, opinions, and local zones. 1. Amoeba Complex Algebraic Varieties: For an algebraic variety $V$ in complex number space $\mathbb{C}^n$, the amoeba $\mathcal{A}(V)$ is defined as follows:

$$\mathcal{A}(V) = \{(\log|z_1|, \log|z_2|, \ldots, \log|z_n|) \mid (z_1, z_2, \ldots, z_n) \in V\}$$

where $z_i$ are complex coordinates on $V$.

2. Regarding Ronkin Functions in This Thesis: Ronkin functions are a concept in tropical geometry and are closely related to amoeba theory. Ronkin functions calculate the "weights" of each connected component of the complement of the amoeba. For the amoeba $\mathcal{A}(V)$ of a complex algebraic variety $V$, Ronkin functions are assigned to each connected component of its complement $\mathbb{R}^2 \setminus \mathcal{A}(V)$.

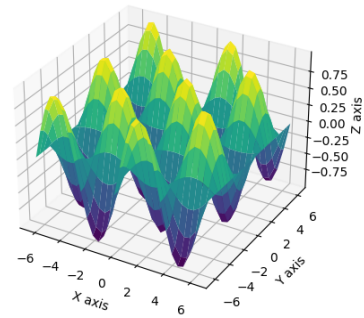

Fig. 8: Surface Representation:Monge Ampere equation, Amoeba and Ronquin numbers:1

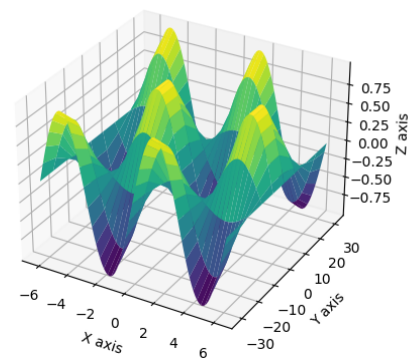

Fig. 9: Surface Representation:Monge Ampere equation, Amoeba and Ronquin numbers:2

3. Periodic Hexagonal Lattice: A periodic hexagonal lattice refers to a pattern of regularly spaced equilateral hexagons in the plane. In this context, it is assumed that the connected components of the complement of the amoeba $\mathcal{A}(V)$ correspond to such periodic hexagonal lattices.

4. Strict Convexity of the Complement of the Amoeba: The strict convexity of the complement of the amoeba means that its connected components are convex sets. In other words, it describes the property that any line segment connecting any two points in the complement of the amoeba is completely contained within the complement.

We will clarify the definition concerning the strict convexity of the complement of the amoeba. It refers to the convexity of the complement of the amoeba, combining concepts from algebraic geometry and convex analysis.

### 3.0.1 Definition of Convex Sets

A convex set is a set where any line segment between any two points within the set is completely contained within the set itself. Mathematically, a set $C \subset \mathbb{R}^n$ is convex if, for any $x, y \in C$ and any $\lambda$ such that $0 \leq \lambda \leq 1$, the following equation holds:

$$\lambda x + (1 - \lambda) y \in C$$

### 3.0.2 Amoebas and Their Complement

An amoeba is the image of a complex algebraic variety under the logarithmic absolute value map. Specifically, for an algebraic variety $V$ in complex number space $\mathbb{C}^n$, the amoeba $\mathcal{A}(V)$ is defined as follows:

$$\mathcal{A}(V) = \{(\log|z_1|, \log|z_2|, \ldots, \log|z_n|) \mid (z_1, z_2, \ldots, z_n) \in V\}$$

The complement of the amoeba is $\mathbb{R}^n \setminus \mathcal{A}(V)$, which refers to the region in $\mathbb{R}^n$ not occupied by the amoeba.

### 3.0.3 Strict Convexity

The strict convexity of the complement of the amoeba means that this complement has convexity. In other words, for any two points $p, q$ in the complement, the line segment connecting them is entirely contained within the complement. Mathematically, if $p, q \in \mathbb{R}^n \setminus \mathcal{A}(V)$, then for any $0 \leq \lambda \leq 1$, the following holds:

$$\lambda p + (1 - \lambda) q \in \mathbb{R}^n minus \mathcal{A}(V)$$

### 3.0.4 Significance of Convexity of the Complement of the Amoeba

The convexity of the complement of the amoeba is crucial in tropical geometry and algebraic geometry. It suggests that the shape and structure of the complement of the amoeba are closely related to the algebraic properties of the variety $V$. Studying the convexity of the complement of the amoeba provides various profound insights into the variety. In particular, the convexity of the complement of the amoeba often arises when the variety is defined by specific types of algebraic equations.

### 3.0.5 The Convexity of the Amoeba's Complement and Its Stereoscopic Understanding in the Periodic Hexagonal Lattice Dimer Model

This problem lies at the intersection of combinatorics and statistical physics.

### 3.0.6 Consider a Bipartite Graph and Dimer Model on the Torus

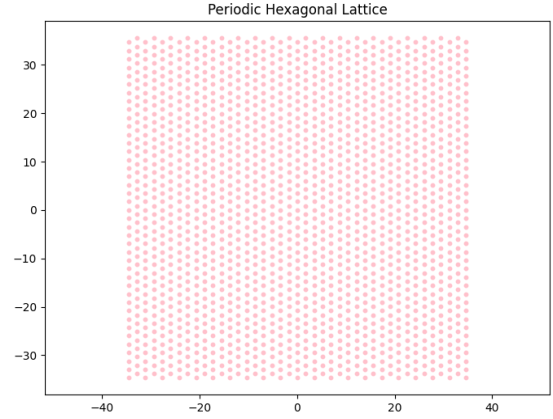

Fig. 10: Surface Representation:Periodic Hexagonal Lattice $s = 10$

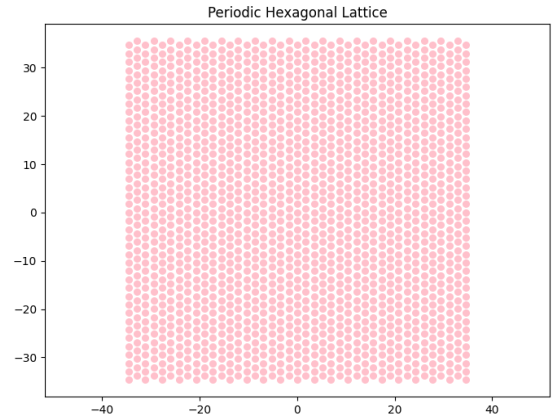

Fig. 11: Surface Representation:Periodic Hexagonal Lattice $s = 30$

The periodic hexagonal lattice consists of hexagons, with each vertex connected by three edges. By mapping this lattice onto a torus, it can be treated as an infinite lattice without boundaries. In the dimer model, these edges are covered by "dimers" (pairs of connected edges).

In a bipartite graph on the torus, each vertex of the lattice is colored either "black" or "white," and vertices of different colors are connected. Dimers are represented as edges connecting vertices of different colors.

Now, introduce the Kasteleyn matrix. It is used to calculate the number of dimer configurations on a bipartite graph on the torus. This matrix is constructed as follows:

1. Associate "black" and "white" vertices of the graph with rows and columns, respectively. 2. If an edge $e$ of the graph connects "black" vertex $i$ and "white" vertex $j$, assign a nonzero value (e.g., ±1) to the matrix element $(i, j)$. This value depends on the orientation of the edge and the specific structure of the graph.

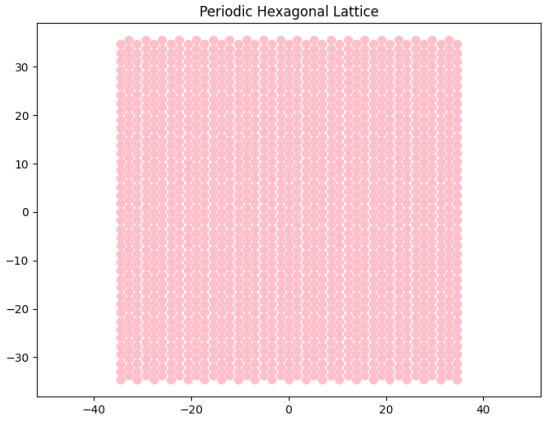

Fig. 12: Surface Representation:Periodic Hexagonal Lattice $s = 50$

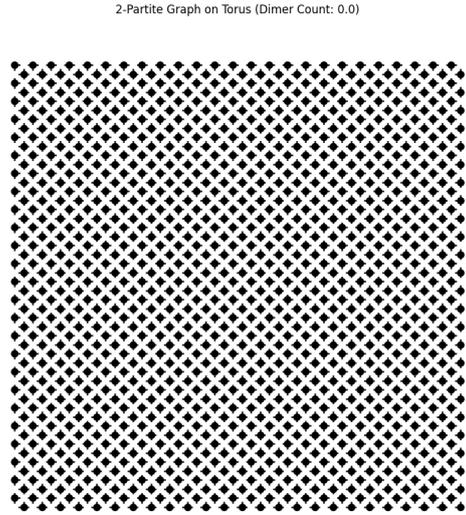

Fig. 14: 2-Partite Graph on Torus $n = 50$

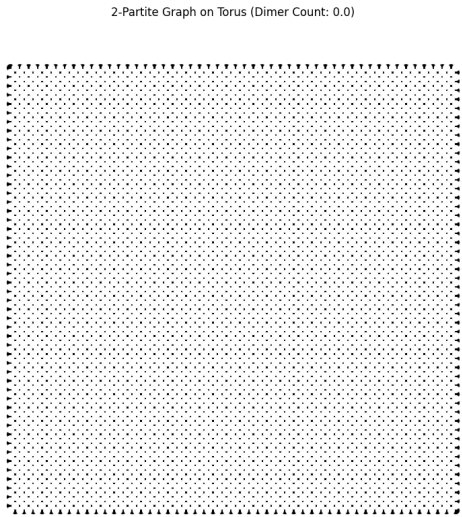

Fig. 13: 2-Partite Graph on Torus $n = 100$

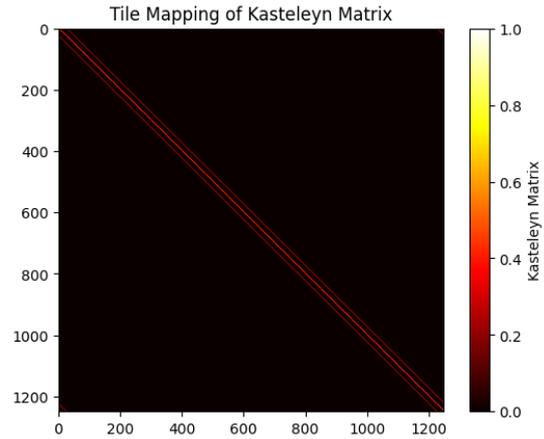

Fig. 15: Tile Mapping of Kasteleyn Matrix

3. The total number of dimer configurations is given by the determinant (or its absolute value) of this Kasteleyn matrix.

To connect the convexity of the amoeba's complement with the dimer model, it is assumed that the complement of the amoeba corresponds to a manifold with certain algebraic properties, which can be mapped to the dimer configuration problem.

To establish the connection between the convexity of the amoeba's complement and the dimer model, it is necessary to first mathematically understand each concept and then consider how to map them. Here, we provide a basic approach with mathematical notation.

The amoeba manifold is the image of a complex algebraic variety $V \subset \mathbb{C}^n$ under the logarithmic absolute value map. The amoeba $\mathcal{A}(V)$ is defined as follows:

$$\mathcal{A}(V) = \{(\log|z_1|, \ldots, \log|z_n|) \in \mathbb{R}^n \mid (z_1, \ldots, z_n) \in V\}$$

The convexity of the complement of the amoeba, $\mathbb{R}^n \setminus \mathcal{A}(V)$, means that for any two points $x$, $y$ in the complement, the line segment $\lambda x + (1 - \lambda)y$ (where $0 \leq \lambda \leq 1$) is entirely contained within the complement.

The dimer model is a problem of random matchings on a graph. In a graph $G$, a dimer configuration is a set of edges where each vertex is connected to exactly one edge.

In the periodic hexagonal lattice dimer model, each dimer covers an edge of the hexagonal lattice. In a bipartite graph $G$ on the torus, the Kasteleyn matrix $K$ is used to calculate the number of dimer configurations.

Points within the complement of the amoeba manifold are related to the possibilities of dimer configurations. This correspondence can be seen as points in the complement representing the "energy" or "probability weight" of dimer configurations.

### 3.0.7 The Relationship Between the Convexity of the Amoeba's Complement and the Stochastic Weighting and Arrangement in the Dimer Model

**The Convexity of the Amoeba's Complement**

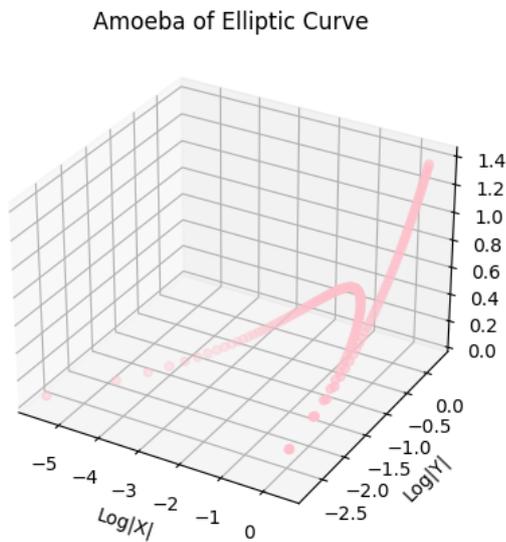

Fig. 16: Surface Representation:Amoeba of Elliptic Curve

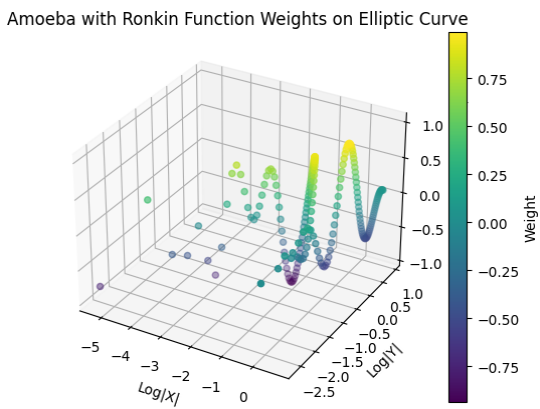

Fig. 17: Surface Representation:Amoeba with Ronkin Function Weights on Elliptic Curve

The convexity of the amoeba's complement implies that any straight line segment between any two points within the complement is entirely contained within the complement. Mathematically, when the complement of the amoeba, $\mathcal{A}^c$, is convex, the following relation holds for any $x, y \in \mathcal{A}^c$ and $0 \leq \lambda \leq 1$:

$$\lambda x + (1 - \lambda) y \in \mathcal{A}^c$$

### 3.0.8 The Amoeba's Complement and Its Properties

Consider the region within the amoeba $\mathcal{A}(V)$ and denote it as $\mathbb{R}^2 \setminus \mathcal{A}(V)$. Specific points within the complement of the amoeba $x \in \mathbb{R}^2 \setminus \mathcal{A}(V)$ are believed to correspond to specific algebraic properties of the manifold $V$.

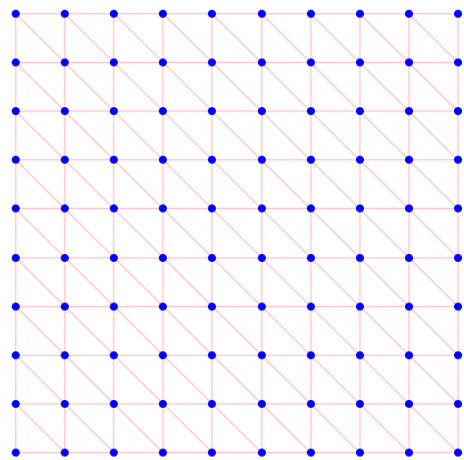

Fig. 18: Hexagonal Lattice Network Graph

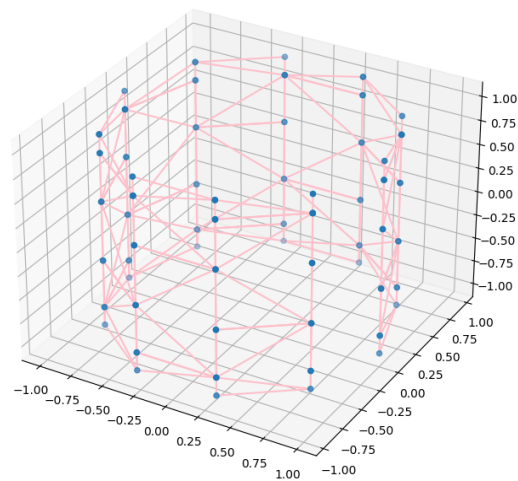

Fig. 19: Surface Representation:Torus Graph of Hexagonal Lattice

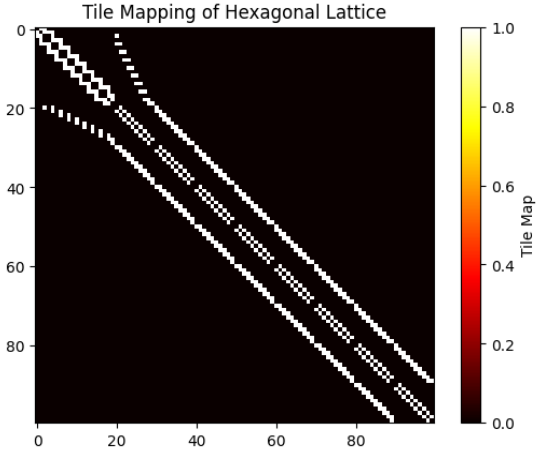

Fig. 20: Tile Mapping of Hexagonal Lattice

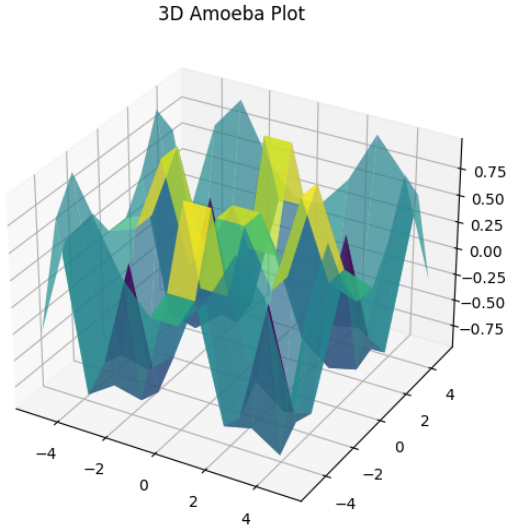

Fig. 21: Surface Representation:Amoeba Map

#### 3.0.9 Stochastic Model of Dimer Configurations

Let's consider a stochastic model of dimer configurations. The probability of a dimer existing on each edge $e$ of a graph $G$ is $p_e$, and the probability of a complete dimer configuration is $P(D)$. Here, $D$ represents a specific dimer configuration on the graph $G$, and this probability may be expressed as follows:

$$P(D) = \frac{1}{Z} \prod_{e \in D} p_e$$

Here, $Z$ is the normalization constant (partition function).

#### 3.0.10 Associating the Amoeba's Complement with Dimer Configurations

Points $x$ within the complement of the amoeba are considered to influence the stochastic weighting of dimer configurations. For example, point $x$ may determine the "energy" of the dimer configuration, which can affect the probability $p_e$. This may be modeled as follows:

$$p_e = \exp(-\beta E(x, e))$$

Here, $E(x, e)$ is the energy function, and $\beta$ is the inverse temperature parameter. This energy function represents the relationship between points $x$ within the complement of the amoeba and the edge $e$, and the specific definition of this function depends on the particular physical or mathematical model.

Ultimately, the probability $P(D)$ of dimer configurations can be constructed as a model that depends on the properties of points $x$ within the complement of the amoeba:

$$P(D) = \frac{1}{Z} \prod_{e \in D} \exp(-\beta E(x, e))$$

This model suggests how the geometric and algebraic properties of the amoeba's complement may influence the statistical properties of dimer configurations.

#### 3.0.11 Quantification of Dimer Configuration Probabilistic Weights Based on the Convexity of Amoeba Complements

**1. Definition of the Energy Function**

We define the energy function $E(x, e)$ based on the convex region within the complement of the amoeba, where $x$ is a point in the complement and $e$ is an edge of a dimer. Points $x$ within the convex region can be considered to reduce the energy of dimer configurations. For example, the energy function might be defined as:

$$E(x, e) = -\alpha \cdot \text{dist}(x, \text{ConvexRegion})$$

Here, $\text{dist}(x, \text{ConvexRegion})$ represents the distance from point $x$ to the nearest convex region, and $\alpha$ is a positive coefficient.

**2. Probability of Dimer Configurations**

The probability model for dimer configurations is as follows:

$$P(D) = \frac{1}{Z} \exp\left(-\beta \sum_{e \in D} E(x, e)\right)$$

Here, $Z$ is the normalization constant (partition function), and $\beta$ is the inverse temperature parameter.

## 3. Calculation of the Partition Function

The partition function $Z$ is the sum of $P(D)$ over all possible dimer configurations:

$$Z = \sum_D \exp\left(-\beta \sum_{e \in D} E(x, e)\right)$$

## 4. Analysis of the Relationship Between Convex Regions and Dimer Configurations

We observe how the probability distribution of dimer configurations changes by altering the position or shape of the convex region. This involves changing the convex region and recalculating the partition function $Z$ and each $P(D)$.

We describe specific solutions for calculating the partition function using Monte Carlo methods and matrix-tree theorem. First, we explain the basic ideas of each method, and then we show how to apply these to the calculation of the partition function.

**Monte Carlo Method**

The Monte Carlo method uses probabilistic sampling for numerical analysis. In this case, we generate random samples from the probability distribution of dimer configurations and use these to approximate the partition function.

**Formulas and Solution**

1. **Initialization**: Select an appropriate initial dimer configuration $D_0$. 2. **Iteration**: At each step, generate a new configuration $D_{i+1}$ randomly from the current configuration $D_i$. This generation process might involve, for example, flipping a randomly chosen edge. 3. **Probability Calculation**: Calculate $P(D_i)$ for each dimer configuration $D_i$ and use this to obtain an approximate value for the partition function $Z$. For example,

$$Z \approx \frac{1}{N} \sum_{i=1}^{N} \exp(-\beta \sum_{e \in D_i} E(x, e))$$

Here, $N$ is the number of samples.

**Matrix-Tree Theorem**

The matrix-tree theorem is used to calculate the number of all spanning trees of a graph. In the dimer model, it can be used to calculate the number of dimer configurations using the determinant of the Kasteleyn matrix.

**Formulas and Solution**

1. **Construction of the Kasteleyn Matrix**: Construct the Kasteleyn matrix $K$ for the graph $G$. This matrix is similar

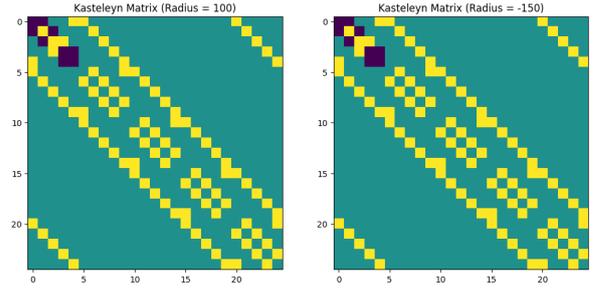

Fig. 22: Kasteleyn Matrix

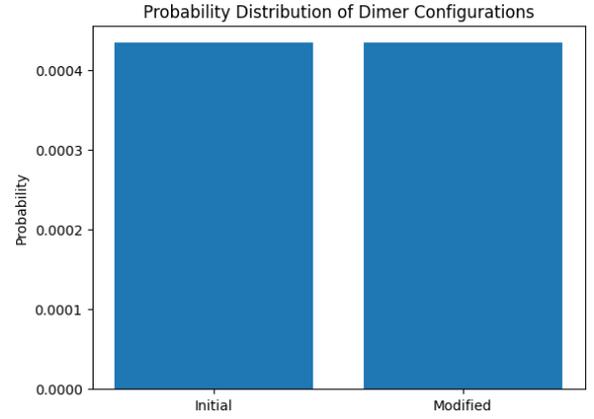

Fig. 23: Probability Distribution of Dimer Configurations

to the adjacency matrix of the graph but considers the orientation and energy of dimer configurations. 2. **Calculation of the Determinant**: Calculate the determinant of $K$. According to the matrix-tree theorem, the absolute value of this determinant equals the number of spanning trees in the graph. 3. **Calculation of the Partition Function**: The number of dimer configurations is proportional to the absolute value of the determinant, so the partition function $Z$ is calculated as

$$Z = |\det(K)|$$

The Monte Carlo method is suitable for approximating the partition function based on sampling from the probability distribution, but it may require a long computation time. On the other hand, the matrix-tree theorem is efficient for certain types of dimer configurations, but it is not applicable to all dimer configurations. These methods will be chosen depending on the nature of the problem and computational resources available.

### 3.0.12 Concrete Computational Example for Analyzing the Relationship Between Convex Regions and Dimer Configurations

Define a convex region $C$ within the complement of the amoeba and set up an energy function $E(x, e)$ for points $x$

within this region. It is assumed that points within the convex region reduce the energy of dimer configurations. For example, you can set it up as follows:
- For points $x$ within the convex region $C$, the energy function is $E(x, e) = -\alpha$ ($\alpha$ is a positive constant). - For points $x$ outside the convex region, the energy function is $E(x, e) = 0$.

### 3.0.13 Calculation of the Partition Function

Calculate the partition function $Z$ and the probability $P(D)$ for individual dimer configurations $D$:
- $Z = \sum_D \exp(-\beta \sum_{e \in D} E(x, e))$ - For each dimer configuration $D$, $P(D) = \frac{1}{Z} \exp(-\beta \sum_{e \in D} E(x, e))$

### 3.0.14 Alteration of Convex Region and Analysis of Probability Distribution

Next, change the position and shape of the convex region $C$ and recalculate both the partition function $Z$ and each $P(D)$ accordingly. For example, you can move the convex region to a different location or modify its size and shape to observe how the probability distribution of dimer configurations changes in response.

### 3.0.15 Specific Computational Example

As an example, consider changing the convex region $C$ from a small circle to a large circle. This alteration will affect the energy functions $E(x, e)$ associated with points within the convex region, thereby causing changes in the values of the partition function $Z$ and each $P(D)$.

(1)Calculate the partition function and probabilities for the small circular convex region. (2)Change to a large circular convex region and calculate the partition function and probabilities again. (3)Compare and analyze the changes in the probability distributions in both cases.

Through this comparison, you can quantitatively understand how the size and position of the convex region affect the probabilities of dimer configurations.

### 3.0.16 Impact of Amoeba Complement Convexity on Dimer Models

Understanding the Impact of Amoeba Complement Convexity on Dimer Models: Formulas and Solutions for Comparatively Analyzing Probability Distributions for Different Convex Regions

### 1. Definition of Energy Function

Define an energy function $E(x, e)$ based on the convex region $C$ within the complement of the amoeba. This function should take low values (e.g., $E(x, e) = -\alpha$) for points $x$ inside the convex region and high values (e.g., $E(x, e) = 0$) for points $x$ outside the convex region.

### 2. Calculation of the Partition Function and Dimer Configuration Probabilities

Calculate the partition function $Z$ and the probabilities $P(D)$ for individual dimer configurations $D$ as follows:
- Partition function: $Z = \sum_D \exp(-\beta \sum_{e \in D} E(x, e))$
- Probability for each dimer configuration: $P(D) = \frac{1}{Z} \exp(-\beta \sum_{e \in D} E(x, e))$

### 3. Alteration of Convex Region and Analysis of Probability Distributions

Change the position and shape of the convex region $C$ and recalculate the partition function $Z$ and each $P(D)$ accordingly. Calculate probability distributions for dimer configurations for different convex regions and compare the changes in these distributions.

**Specific Solution**

**Convex Region Setup**

Set up the initial convex region $C_1$ and calculate the energy function $E(x, e)$ for points within it.

**Calculation of Initial Distribution**

Calculate $Z$ and $P(D)$ based on $C_1$.

**Change in Convex Region**

Modify the convex region to $C_2$ and calculate the energy function $E(x, e)$ again.

**Calculation of Updated Distribution**

Based on $C_2$, calculate the new $Z$ and $P(D)$.

**Comparative Analysis**

Compare the probability distributions $P(D)$ under $C_1$ and $C_2$, and analyze how the change in the convex region affects the probabilities of dimer configurations.

This analysis allows for a quantitative understanding of how the convex regions within the amoeba complement influence the statistical physical properties of dimer models.

To understand the impact of amoeba complement convexity on dimer models, the idea of modeling social dependence areas such as "digital dependence" and "cognitive forgetting" is an abstract and creative approach. Here, we consider local opinion dynamics with dependencies and forgetfulness using dimer configurations.

## 1. Dimer Model

Consider dimer configurations on a graph where each edge represents specific social opinions or behaviors (e.g., smartphone dependence).

## 2. Definition of Dependency Region

Within the complement of the amoeba, model a specific convex region as a "dependency region." Points within this region are assumed to strongly represent social dependence.

**Energy Function and Kirchhoff Matrix**

### 1. Energy Function

Assign low energy to dimer configurations within the dependency region and high energy to configurations outside the region.

### 2. Updating Kirchhoff Matrix

Update elements of the Kirchhoff matrix based on the dependency region, assigning high weight (low energy) to edges within the dependency region and low weight (high energy) to edges outside the region.

**Partition Function and Probability of Dimer Configurations**

### 1. Calculation of Partition Function

The partition function $Z$ is calculated based on the absolute value of the determinant of the Kirchhoff matrix.

### 2. Calculation of Probabilities

The probability $P(D)$ for each dimer configuration is calculated using $Z$:

$$P(D) = \frac{\exp(-\beta \sum_{e \in D} E(x,e))}{Z}$$

# 4. Modeling Local Opinion Dynamics

## 1. Parameters for Dependency and Forgetfulness

Introduce parameters $\alpha$ (dependency strength) to represent dependence and $\gamma$ (forgetting rate) to represent forgetfulness.

## 2. Dynamics Calculation

Using the dependency and forgetfulness parameters, model the probabilistic change of dimer configurations over time. This can involve stochastic simulation methods like Markov Chain Monte Carlo, among others.

The energy function can be set as follows, for example:

$$E(x,e) = \begin{cases} -\alpha & \text{if } x \in \text{Dependency Region} \\ \gamma & \text{otherwise} \end{cases}$$

Here, $x$ represents vertices on the graph, and $e$ represents dimer configurations. Such a model is highly intriguing for mathematical modeling of social phenomena and behavioral patterns. However, capturing the dynamics of real social dependence may require incorporating more complex factors and phenomena into the model. Additionally, the results of this model may not directly apply to real social behaviors and opinion formation, but they have the potential to provide theoretical insights. To analyze the variation in the probability distribution of dimer configurations for different convex regions and understand the impact of amoeba complement convexity on the dimer model, we employ the use of the Kirchhoff matrix in mathematical expressions.

## Concept of the Kirchhoff Matrix

The Kirchhoff matrix is used to calculate the number of dimer configurations in the dimer model. It is a matrix with entries corresponding to each edge in the graph, and these entries represent the weight (or energy) of dimer configurations.

## Definition of the Energy Function

Based on the convex regions within the complement of the amoeba, we define an energy function $E(x,e)$. We assign low energy values to edges within the convex region and high energy values to edges outside the convex region.

## Recalculation of the Kirchhoff Matrix

The Kirchhoff matrix is recalculated for different convex regions. The entries of the Kirchhoff matrix $K$ are updated based on the energy function $E(x,e)$. The determinant (or absolute value thereof) of this matrix represents the number of possible dimer configurations on the graph.

## Calculation of the Partition Function and Probabilities

The probability $P(D)$ for dimer configurations is calculated using the partition function $Z$. This is expressed as:

$$Z = \sum_D \exp(-\beta \sum_{e \in D} E(x,e))$$

$$P(D) = \frac{\exp(-\beta \sum_{e \in D} E(x,e))}{Z}$$

Here, $\beta$ is the inverse temperature parameter, and $D$ represents a specific dimer configuration.

## Analyzing the Impact of Convex Region Changes

For different settings of convex regions, we compute the Kirchhoff matrix $K$, the partition function $Z$, and the probabilities of dimer configurations $P(D)$. This allows us to analyze how changes in convex regions affect the probabilities of dimer configurations.

## Interpretation of Results

By comparing the changes in probability distributions $P(D)$ for different convex region settings, we can quantitatively understand how amoeba complement convexity impacts the statistical physics properties of the dimer model. This analysis can reveal the influence of specific arrangements or shapes of convex regions on the probability distribution of the dimer model.

Mapping the Results of Opinion Dynamics to Amoeba Complement Convexity and the Dimer Model

### 4.1 Mapping Amoeba Complement Convexity to the Dimer Model

### 1. Results of Opinion Dynamics

Let the results of opinion dynamics be denoted as $O(x, t)$, where $x$ represents positions on the graph, and $t$ represents time.

### 2. Mapping to Amoeba Complement

Define the convexity at points $y$ within the amoeba's complement based on the results of opinion dynamics $O(x, t)$. For example, the strength of opinions at specific positions in the convex region may be considered to represent the degree of convexity at those points.

### 4.2 Ronkin Functions and Legendre Transform

The Ronkin function $L(x)$ calculates the "weight" of connected components at points $x$ within the amoeba's complement. Legendre transforms are typically used to investigate the properties of convex functions, but directly performing a Legendre transform on Ronkin functions is not common. However, as a theoretical approach, it can be considered as follows:

### 1. Ronkin Function

$L(x)$**: Consider the Ronkin function at points $x$ within the amoeba's complement.

### 2. Legendre Transform

The Legendre transform of the Ronkin function $\Phi(p)$ may be defined as follows:

$$\Phi(p) = \sup_{x} \left[ \langle p, x \rangle - L(x) \right]$$

Here, $p$ represents the dual variable, and $\langle p, x \rangle$ denotes the inner product.

This Legendre transform can potentially offer an alternative perspective to capture the characteristics of amoeba complement convexity, viewing it from a different standpoint.

## 5. Discussion

### 5.1 Analysis of Convex Regions in Opinion Dynamics

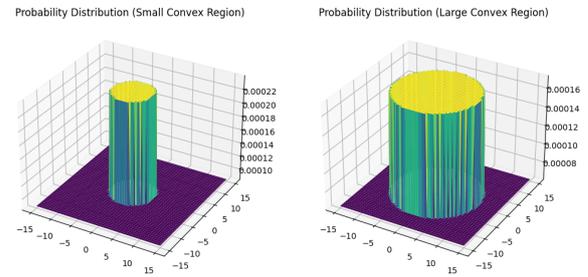

Fig. 24: Probability Distribution (Small, Large Convex Region)

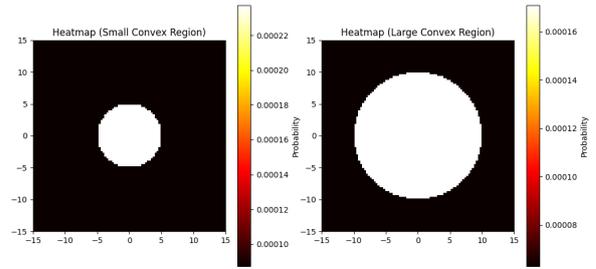

Fig. 25: Heatmap (Small, Large Convex Region)

In this analysis, we explore the effects of different sized convex regions on the probability distributions in an opinion dynamics model. We use a grid-based approach to calculate the energy function and the subsequent probability distribution across two different convex regions.

### 5.2 Formulas and Parameters

The core of our analysis is based on the following parameters and formulas:

Grid size: $n\_points = 100$

Energy coefficients: $\alpha = 1.0$ and $\beta = 1.0$

Radii of convex regions: $radius\_small = 5$ and $radius\_large = 10$

Grid definition: $x, y$ ranges from -15 to 15.

Energy Function:

$$E(x, y, radius) = \begin{cases} -\alpha & \text{if } \sqrt{x^2 + y^2} \leq radius \\ 0 & \text{otherwise} \end{cases}$$

Distribution Function:

$$Z = \sum \exp(-\beta \cdot E)$$

$$P = \frac{\exp(-\beta \cdot E)}{Z}$$

The intent of this analysis is to understand how the size of the convex region impacts the probability distribution in a spatial model. The model simulates opinion dynamics by considering different energy levels within and outside convex regions. Energy is lower inside the convex region, symbolizing a higher level of agreement or consensus. Two scenarios are examined: one with a smaller convex region and another with a larger one. This comparison helps in understanding the role of spatial factors in opinion dynamics and consensus formation.

The Results depict two sets of graphical representations illustrating a probability distribution across a two-dimensional grid. These representations demonstrate how the size of a convex region influences the distribution within the framework of an opinion dynamics model. The first set of images is presented as a 3D cylindrical graph, while the second set corresponds to a 2D heatmap.

The probability distribution is governed by an energy function that relies on the distance between the grid's center and a point $(x, y)$, as well as whether the point falls within a specified convex region defined by its radius. The smaller convex region has a radius of 5 units, while the larger one has a radius of 10 units.

(1) **Social Phenomena Consideration:** The graphics may be illustrating the concentration of opinions within a social network, where opinion dynamics are constrained by geographical or ideological boundaries represented by the convex regions. The smaller region could signify a closely-knit community with a strong consensus, resulting in a higher probability density. In contrast, the larger region may represent a more dispersed community with a weaker consensus, leading to a lower probability density.

(2) **Media Influence Consideration:** If the convex regions symbolize the reach or influence of different media sources, these models might be demonstrating that a smaller media source exerts a more potent yet localized impact on public opinion. Conversely, a larger media source has a broader but less intense influence.

(3) **Consensus Formation Consideration:** Concerning consensus formation, these models may suggest that smaller groups (represented by the small convex region) find it easier to reach consensus, resulting in a higher probability density. In contrast, larger groups (indicated by the large convex region) may exhibit a more diverse range of opinions and a less concentrated consensus.

(4) **Consideration of Reconstruction from a Dimer Model to a Torus:** If we interpret the convex regions as constraints within a dimer model, the images could represent the impact of these constraints on the probability distribution of dimers within a torus-shaped lattice. A smaller constraint might correspond to a higher binding energy within a confined area, while a larger constraint signifies a more broadly distributed energy distribution.

(5) **Dimer Model Consideration:** In a dimer model, where each dimer covers two adjacent lattice points, these graphs could represent the likelihood of dimer placement. The smaller region suggests a higher concentration of dimers, implying greater energetic favorability for dimer formation within a constrained space.

(6) **Consideration of Phase Transition of Partial Opinions:** The concept of phase transitions could apply to opinion dynamics, where the high probability density within the small convex region might indicate an ordered phase (consensus), while the lower density within the large region could signify a disordered phase (diverse opinions). Investigating the transition between these phases could involve varying the size of the convex region.

## 5.3 Analysis of Probability Distribution in a Grid-Based Model

This results presents an analytical overview of a grid-based model used for calculating and visualizing probability distributions. The model incorporates the concept of energy functions within a convex region and utilizes network graphing to represent the distribution.

## 5.4 Formulas and Parameters

The model is defined by the following parameters and formulas: Results of four images offers additional visual representations of a probability distribution derived from a grid-based model utilizing an energy function, as defined by the parameters and formulas provided. These visualizations likely correspond to a single convex region with a radius of 5 units.

Grid size: $n\_points = 20$

Energy coefficients: α = 1.0

Inverse temperature parameter: β = 1.0

Radius of the convex region: *radius* = 5

Grid coordinates: *x*, *y* range from -10 to 10.

Also Energy Function:

$$E(x, y) = \begin{cases} -\alpha & \text{if } \sqrt{x^2 + y^2} \leq radius \\ 0 & \text{otherwise} \end{cases}$$

Also Distribution Function:

$$Z = \sum \exp(-\beta \cdot E)$$

$$P = \frac{\exp(-\beta \cdot E)}{Z}$$

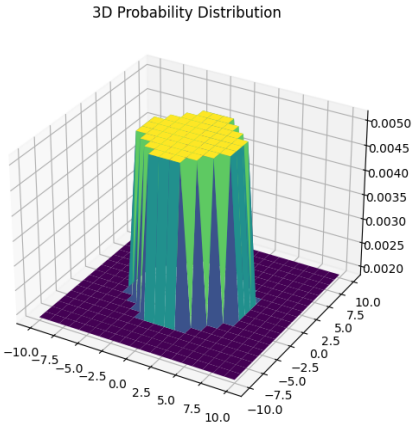

Fig. 26: Probability Distribution

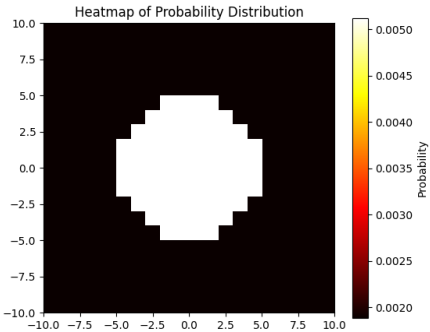

Fig. 27: Heatmap of Probability Distribution

(1) **Social Phenomena Consideration:** The 3D probability distribution and its corresponding visual representations may depict how individuals within a society (as defined by the convex region) are influenced to adopt a common opinion. The concentration of higher probabilities within the convex region could suggest a strong localized agreement or adherence to a cultural norm.

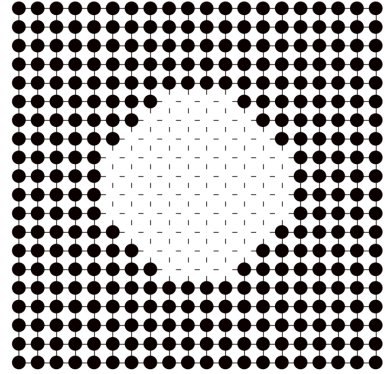

Fig. 28: Network Graph of Probability Distribution

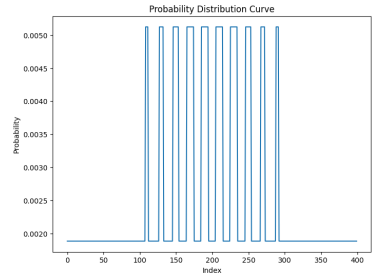

Fig. 29: Probability Distribution Curve

(2) **Media Influence Consideration:** These visualizations could serve as a means to examine the impact of media within a defined range (the convex region). The uniformity observed within the convex region may indicate that media influence is substantial and evenly spread within that area but does not extend beyond its boundaries.

(3) **Consensus Formation Consideration:** The network graph may represent individuals within a network, while the heatmap illustrates the likelihood of consensus formation within the community. The distinct boundary of the convex region implies that consensus is more likely to be achieved within this community than outside of it.

(4) **Reconstruction from a Dimer Model to a Torus Consideration:** In the context of a network of dimers, the uniform distribution within the convex region on the network graph might suggest a stable configuration of dimers. This could be particularly relevant when considering the transition of a dimer model from a planar to a toroidal topology.

(5) **Dimer Model Consideration:** The network graph might showcase potential sites for dimers in a lattice model, where the central area (convex region) indicates a high probability of dimer occupation due to favorable energy states.

(6) **Phase Transition of Partial Opinions Consideration:**
The probability distribution could be employed to model phase transitions within a system of opinions. The sharp boundary of the convex region in the heatmap could signify a critical threshold, where a phase transition occurs, distinguishing a phase of uniform opinions within the region from a phase of diverse opinions outside of it.

These results representations are likely generated by assigning lower energy values (resulting in higher probabilities) to grid points situated within the convex region, as dictated by the provided energy function. The distribution function subsequently normalizes these probabilities across the entire grid.

# 6. Analysis of Opinion Dynamics on a Torus Graph

This results details the simulation parameters, analytical intent, and visualization techniques for a study of opinion dynamics on a torus graph. The aim is to study how opinions evolve over time on a torus graph, considering the effects of dependence and forgetting. The convex region represents an area of high consensus or agreement, and its impact on the overall opinion dynamics is analyzed.

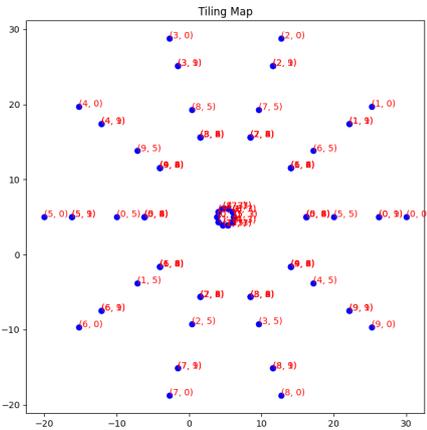

Fig. 30: Opinion Dynamics:Tiling Map

## 6.1 Fitting Parameters

The following parameters define the opinion dynamics model:

Number of grid points: $n\_points = 10$

Number of time steps for simulation: $timesteps = 100$

Dependence strength: $alpha = 0.0$

Forgetting rate: $gamma = 0.05$

Radius of the convex region: $radius = 5$

Torus graph dimensions: $n = n\_points, r = 20, R = 5$

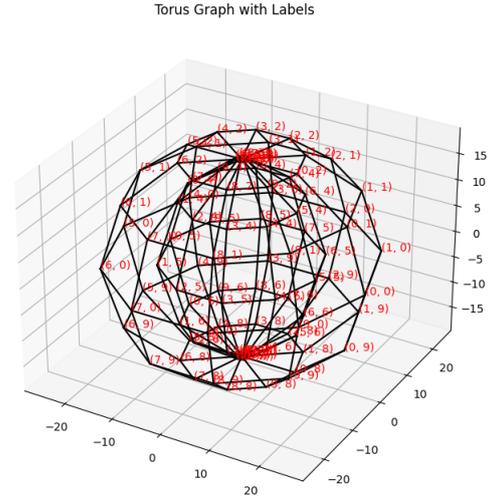

Fig. 31: Opinion Dynamics:Tiling Map, Torus Network

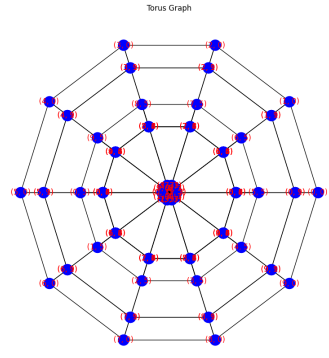

Fig. 32: Opinion Dynamics:Tiling Graph

# 7. Mathematical Formulation

The energy function and opinion update rules are defined as:

Energy Function:

$$E(i, j) = \sqrt{(i - x\_center)^2 + (j - y\_center)^2}$$

Opinion Update:

$$new\_opinions[i, j] = opinions[i, j]+$$

$$\begin{cases} \alpha \cdot (1 - opinions[i, j]), \\ \quad \text{if } E(i, j) < radius \\ -\gamma \cdot opinions[i, j], \\ \quad \text{otherwise} \end{cases}$$

## 7.1 Visualization Techniques

The model employs various visualization methods:

(1) Tiling Map: 2D representation of the torus graph nodes.

(2) **3D Graph:** Visualizes the nodes and edges of the torus graph in a 3D space.
(3) **Network Graph:** Shows the torus graph in 2D with node labels.

This analysis provides insights into the spatial and temporal dynamics of opinions on complex network structures, contributing to the understanding of social dynamics in structured populations.

The Results portray a torus graph with labeled nodes and the mapping of these nodes onto a flat grid. The accompanying description suggests a simulation of opinion dynamics on this torus, incorporating parameters like dependence strength and forgetting rate. These dynamics are subsequently visualized on the torus graph.

(1) **Social Phenomena Consideration:** The distribution of opinions on a torus might symbolize a society with periodic boundary conditions, where the society's edges connect seamlessly to form a closed community with cyclical social dynamics.
(2) **Media Influence Consideration:** The torus graph could simulate media influence that spans the entire society but exhibits a cyclic nature. Dependence and forgetting factors might reflect how older media narratives can resurface and impact current opinions.
(3) **Consensus Formation Consideration:** The model could illustrate how consensus emerges in a society without clear "edges" or "boundaries," where each opinion has the potential to circulate and influence the entire network. The forgetting rate may represent the dissipation of consensus over time, while dependence strength could signify the pull of a strong societal norm.
(4) **Reconstruction from a Dimer Model to a Torus Consideration:** If we interpret the torus as a dimer model, the visualization might showcase how dimer placements (pairs of opinions) interact across a periodic boundary condition. These dynamics may elucidate the attainment of stable configurations over time.
(5) **Dimer Model Consideration:** In a dimer model, each node could symbolize a molecule, and the edges might represent bonds. The torus topology then reveals how molecules at opposite ends can still be neighbors, influencing the overall system's stability.
(6) **Phase Transition of Partial Opinions Consideration:** The opinion dynamics might simulate a phase transition wherein the system shifts from a disordered state (diverse opinions) to an ordered state (consensus) or vice versa. The convex region could represent a 'nucleation' site where this transformation initiates.
(7) **Dependence Consideration:** The dependence strength ($\alpha$) in the model may signify social reliance or peer pressure within the network. A zero dependence value implies that opinions change independently of others, possibly indicating a highly individualistic society.
(8) **Forgetting Consideration:** The forgetting rate ($\gamma$) could indicate the pace at which society forgets past opinions or events, resulting in a gradual loss of memory.

The simulation parameters and mathematical formulations provided indicate a model in which each node updates its opinion based on its current state, the influence of consensus within the convex region, and a natural tendency to forget or move away from previous opinions. The interplay between these factors determines how opinions evolve over time on the torus graph.

## 7.2 Opinion Distribution

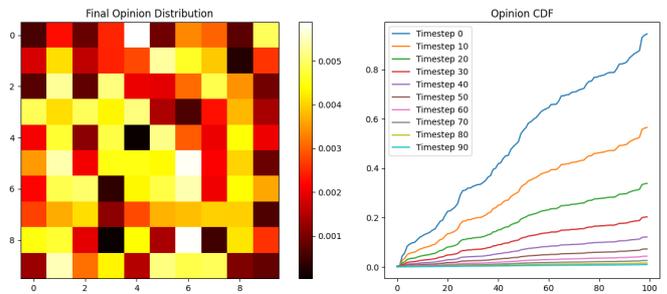

Fig. 33: Opinion Distribution

The provided results conducts a simulation of opinion dynamics over a grid, updating opinions at each timestep based on a rule that considers a central convex region of influence and a forgetting rate.

(a) **Social Phenomena Consideration:** The simulation may replicate how opinions are distributed within a society and how they evolve. The central convex region could symbolize an area of robust social, cultural, or political influence that significantly shapes opinions within its boundaries.
(b) **Media Influence Consideration:** In this context, the media might be construed as a central force influencing opinions. The evolution of the opinion CDF over time may signify the changing impact of media narratives on societal opinions, illustrating how opinions can become more or less diverse as time progresses.
(c) **Consensus Formation Consideration:** The heatmap of the final opinion distribution can reveal regions of high and low consensus. In the model, consensus formation could be impacted by the central region, where opinions are more homogeneous due to stronger local interactions or persuasive communication.

(d) **Reconstruction from a Dimer Model to a Torus Consideration:** Although this simulation isn't directly tied to a dimer model, if we were to interpret it within such a context, the nodes within the convex region might represent areas with a high concentration of dimers, influencing the configuration of the entire network.

(e) **Dimer Model Consideration:** If we envisage the grid nodes as molecules in a dimer model, the simulation could elucidate how molecular interactions lead to intricate patterns of dimer formation, particularly influenced by a central region (potentially representing an area with distinct physical or chemical properties).

(f) **Phase Transition of Partial Opinions Consideration:** The simulation might be employed to explore phase transitions in opinion dynamics, where the system transitions from a disordered state (diverse opinions) to an ordered state (consensus) over time. This transition could be influenced by central 'nucleation' points.

(g) **Dependence Consideration:** Given the 'alpha' parameter is set to 0.0, this simulation assumes no dependence among opinions, implying a highly individualistic society or a model where individuals change their opinions without regard for others' opinions.

(h) **Forgetting Consideration:** The 'gamma' parameter introduces a forgetting mechanism, illustrating how opinions can diminish over time. This aspect may emulate real-world scenarios where past information or beliefs lose influence, potentially permitting the introduction of new opinions or the resurgence of older ones.

In the final opinion distribution, the diverse colors represent the variety of opinions after 100 timesteps, while the CDF plots elucidate how the probability of lower or higher opinions fluctuates over time, offering insights into the dynamics of opinion formation and alteration.

# 8. Opinion Distribution, Dependence and Forgetting

Results provided, collectively represents a simulation of opinion dynamics where the strength of dependence ($\alpha$) and the rate of forgetting ($\gamma$) are not constants but instead follow normal distributions. This simulation aims to model the evolution of each individual's opinion within a grid space while considering the influence of a central region.

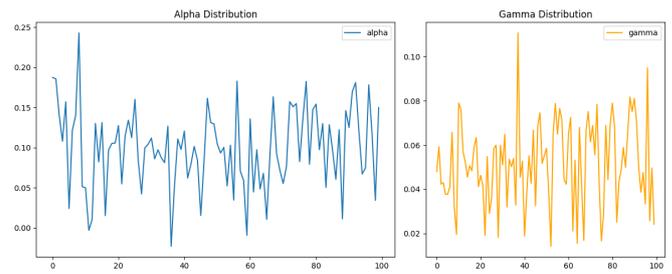

Fig. 34: Opinion Distribution

i. **Phase Transition of Partial Opinions Consideration:** The fluctuating values of $\alpha$ and $\gamma$ can lead to intricate behaviors in opinion dynamics, including phase transitions. For instance, when $\alpha$ is high and $\gamma$ is low at a given timestep, opinions may rapidly converge, resulting in a phase where a single opinion dominates (a consensus phase). Conversely, if $\alpha$ decreases or $\gamma$ increases, opinions might diverge, leading to a disordered phase. The variability in $\alpha$ and $\gamma$ may thus induce transitions between consensus and diversity within the opinion landscape.

ii. **Dependence Consideration:** The distribution of $\alpha$ values across timesteps introduces variability in the strength of dependence among individuals' opinions. A higher $\alpha$ value suggests that individuals are more likely to be influenced by the prevailing opinion within the central region, while a lower $\alpha$ indicates a tendency towards independence. The random variation in $\alpha$ captures the real-world scenario where social influence can fluctuate due to various factors, such as changing societal norms or the impact of significant events.

iii. **Forgetting Consideration:** Similarly, the distribution of $\gamma$ values represents the rate at which individuals forget or move away from their previous opinions. A higher $\gamma$ value suggests a faster rate of opinion change, potentially due to factors like short news cycles or rapidly changing social contexts. The variability in $\gamma$ reflects the dynamic nature of memory and the impact of continuous information flow on opinion stability.

The provided image illustrates the distributions of $\alpha$ and $\gamma$ across the timesteps, providing a visual understanding of how these parameters change over time. This variability could result in a dynamic opinion landscape where periods of high consensus are interspersed with periods of diverse opinions, potentially creating a complex and evolving social model.

## 8.1 Opinion Distribution, Dependence and Forgetting

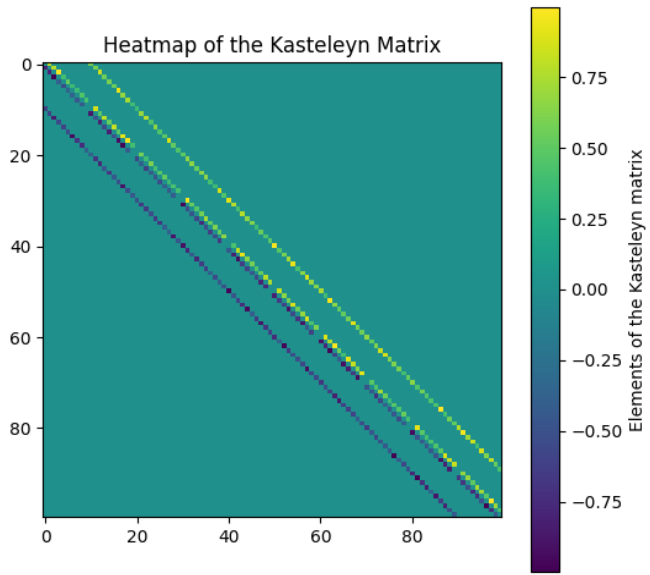

Fig. 35: Heatmap of the Kasteleyn Matrix

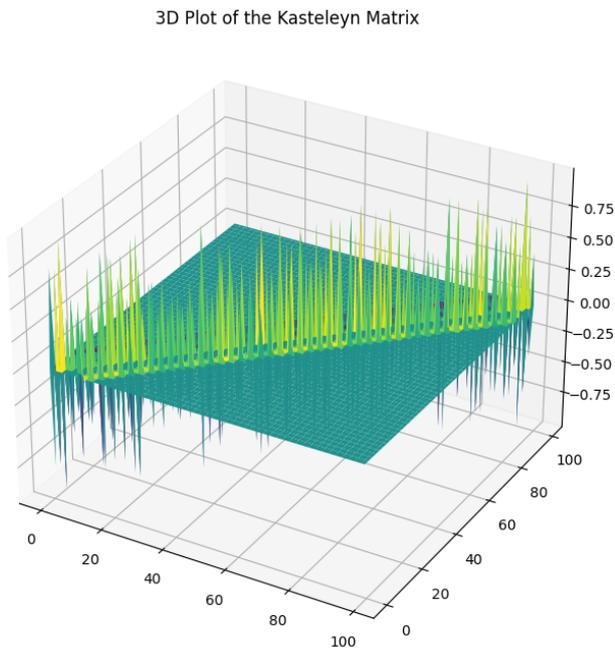

Fig. 36: 3D Heatmap of the Kasteleyn Matrix

Results provided are the outcomes of the supplied

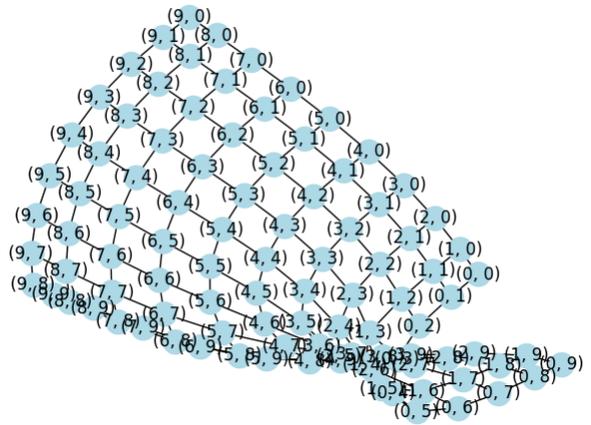

Fig. 37: Network Representation of the Graph

code, which performs a simulation of a physical model using a graph theory approach on a 2D grid.

i. **Social Phenomena Consideration:** The heatmap of the Kasteleyn matrix and the accompanying 3D plot can symbolize the strength of connections within a social network. In this analogy, the grid points represent individuals, and the weights represent the likelihood of interaction or influence. Regions with higher weights (warmer colors) might indicate stronger social ties or a higher probability of interaction, potentially corresponding to social clusters or communities.

ii. **Media Influence Consideration:** If we interpret the graph as a representation of media channels or pathways, the distribution in the Kasteleyn matrix could signify the varying influence that different media outlets exert on different sectors of society. Peaks in the 3D plot might represent areas heavily influenced by media, while valleys might correspond to areas less affected.

iii. **Consensus Formation Consideration:** The network graph depicts the potential pathways through which consensus could be reached in a networked community. The distribution in the Kasteleyn matrix may represent the ease with which consensus can spread through these pathways, influenced by the underlying network topology.

iv. **Reconstruction from a Dimer Model to a Torus Consideration:** Despite the network being represented in a 2D plane, the principles employed in constructing the Kasteleyn matrix could be extended to more intricate topologies like a torus, especially when considering periodic boundary conditions. This extension would alter the enumeration of perfect matchings or consensus states.

v. **Dimer Model Consideration:** The simulation might serve as an analytical tool for studying dimer coverings on a lattice, where the Kasteleyn matrix plays a pivotal role. The determinant of this matrix is closely related to the count of perfect matchings or dimer configurations, which is a fundamental concept in statistical physics.

vi. **Phase Transition of Partial Opinions Consideration:** The variability observed in the Kasteleyn matrix could signify different phases of opinion dynamics within a society. High variation may imply a phase transition, transitioning from a disordered state characterized by diverse opinions to an ordered state where one opinion prevails.

vii. **Dependence Consideration:** The weights assigned to the edges of the graph, as reflected in the Kasteleyn matrix, can be viewed as the strengths of dependence between individuals. The heatmap and 3D plot illustrate the heterogeneity in these dependencies, which may evolve over time as interactions change.

viii. **Forgetting Consideration:** Incorporating forgetting into such models could involve reducing the weights of edges over time or between timesteps. This would result in a dynamic Kasteleyn matrix where the influence of certain interactions decays, simulating the loss of memory or decreasing influence over time.

Results provide an intuitive means of comprehending the abstract concepts inherent in the mathematical model. They help illustrate the intricate interplay between network structure, interaction strengths, and their consequences on the dynamics of the system.

## 8.2 Dependence and Forgetting Matrix

Results simulation of opinion dynamics on a network. The visualization includes a 2D scatter plot showing regions of dependency and forgetting, a 3D surface plot of opinion strengths, and a network representation of a graph with nodes colored based on their region.

i. **Social Phenomena Consideration:** The scatter plot reveals clusters of dependency

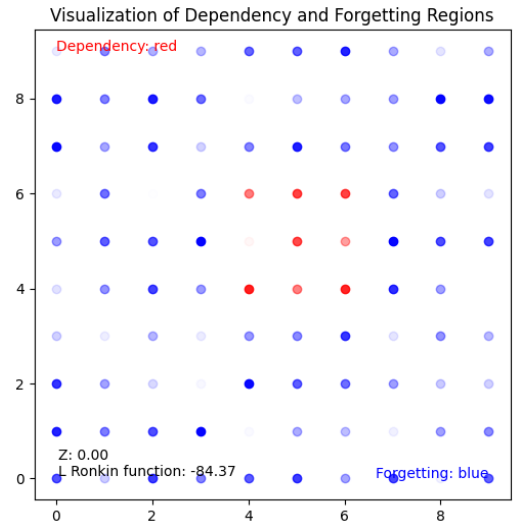

Fig. 38: Dependency and Forgetting Regions Matrix

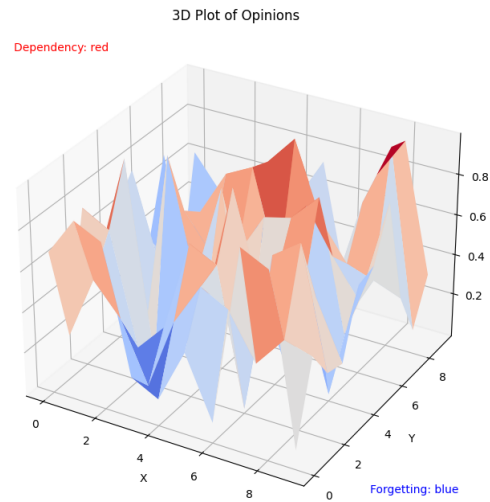

Fig. 39: 3D Dependency and Forgetting Regions Kasteleyn Matrix

(red) and forgetting (blue), implying regions in a social network where opinions are either reinforced or tend to fade away, respectively. This could mirror real-world social structures where specific communities exhibit strong mutual influence, while others experience more fluid opinion dynamics.

ii. **Media Influence Consideration:** The distribution of opinions in the 3D plot might signify how media influence varies across a population. Peaks could represent areas or topics where media narratives currently dominate, while troughs might indicate subjects of lesser focus or older narratives that are being

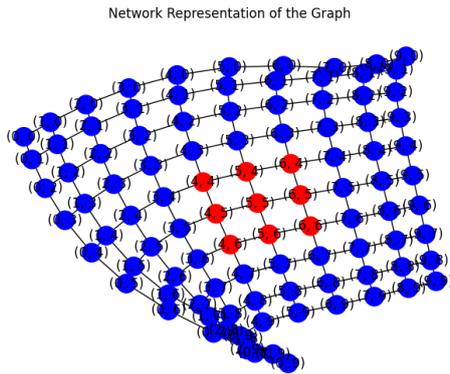

Fig. 40: Network Representation of the Graph:Dependency and Forgetting Regions

forgotten.

iii. **Consensus Formation Consideration:** The network graph featuring red and blue nodes can symbolize how consensus is more likely to form within the red, dependent regions, while the blue regions may denote a diversity of opinions due to a higher rate of forgetting.

iv. **Reconstruction from a Dimer Model to a Torus Consideration:** Although not explicitly shown, the principles demonstrated by the graph can be extrapolated to more complex topologies like a torus. This extrapolation might be useful for studying periodic boundary conditions in opinion dynamics or dimer models.

v. **Dimer Model Consideration:** The Kasteleyn matrix, determinant, and Ronkin function, although not visually depicted, are mathematical constructs that could relate to the enumeration of dimer coverings in a graph. This enumeration method models pair interactions in complex systems.

vi. **Phase Transition of Partial Opinions Consideration:** The variance in opinion strengths within the 3D plot could signify a system undergoing phase transitions, shifting between a diverse set of opinions and a consensus. The transition dynamics depend on the interplay between dependency and forgetting.

vii. **Dependence Consideration:** The red nodes in the network representation might symbolize individuals or groups with strong interdependencies that resist opinion change. In contrast, blue nodes may represent less interdependent entities that are more susceptible to forgetting or altering their opinions.

viii. **Forgetting Consideration:** The blue regions in the scatter and network plots may represent the concept of forgetting in a social network. They indicate individuals or groups that are more inclined to change their opinions, forget past influences, or be influenced by new information.

## 9. Conclusion

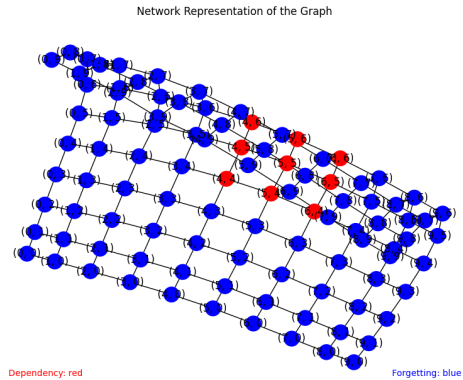

Fig. 41: Dependency and Forgetting Regions Matrix:Network Representation of the Graph

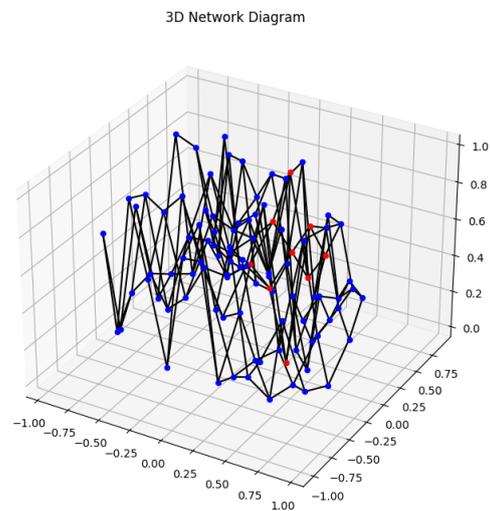

Fig. 42: 3D Dependency and Forgetting Regions Matrix:Network Representation of the Graph

results fgenerates a two-dimensional grid graph and visualizes it in both 2D and 3D forms. Nodes are colored based on their proximity to a defined central area, indicating regions of dependency (red) and forgetting (blue).

A. **Social Phenomena Consideration:** The distribution of red and blue nodes could illustrate how social cohesion (dependency) or dispersion (forgetting) occurs in social networks. Red nodes may represent tightly-knit communities with strong interdependencies, while blue nodes may represent areas where social ties are weaker, leading to a faster turnover of ideas or beliefs.
B. **Media Influence Consideration:** The visualization could be interpreted as the varying degrees of media influence across a population. Red nodes might be areas where media narratives have taken hold and are strongly influencing the population, whereas blue nodes could be less influenced or more quickly moving on to new narratives.
C. **Consensus Formation Consideration:** The network graph may show the potential for consensus-building within a community. Red nodes represent areas where a consensus is more likely to form due to the higher dependency, while blue nodes may represent a diversity of opinions due to the increased rate of forgetting.
D. **Reconstruction from a Dimer Model to a Torus Consideration:** While the network is visualized in 2D and 3D, the principles could be extended to analyze models on a torus or other complex topologies, particularly in considering how local dependencies might affect global consensus on a periodic boundary condition.
E. **Dimer Model Consideration:** The grid graph could represent the underpinnings of a dimer model where red nodes indicate a higher probability of dimer formation (pairing), perhaps due to favorable conditions or energies, while blue nodes indicate areas less conducive to stable dimer formation.
F. **Phase Transition of Partial Opinions Consideration:** The contrasting colors in the visualization could be indicative of a system that is capable of phase transitions, where certain conditions or external influences could push the system from a disordered state (high opinion diversity) to an ordered state (consensus) and vice versa.
G. **Dependence Consideration:** The red nodes are suggestive of areas with strong interdependencies, which could correspond to societal segments where group norms or peer pressure are particularly influential in shaping individual behaviors or beliefs.
H. **Forgetting Consideration:** The blue nodes might represent the transient nature of information or beliefs in certain parts of the network, where ideas can be quickly replaced, and the retention of past opinions is low.

These visualizations serve as a model to understand complex dynamics within networks, highlighting the interplay between local interactions and overall structure. The 3D representation adds depth to the analysis, showing how these dynamics might play out over time or in more complex, multi-layered systems.

# Aknowlegement

The author is grateful for discussion with Prof. Serge Galam and Prof.Akira Ishii. This research is supported by Grant-in-Aid for Scientific Research Project FY 2019-2021, Research Project/Area No. 19K04881, "Construction of a new theory of opinion dynamics that can describe the real picture of society by introducing trust and distrust".